\documentclass[journal]{IEEEtran}
\usepackage[utf8]{inputenc}
\usepackage[T1]{fontenc}
\usepackage{multirow}
\usepackage[table]{xcolor}
\usepackage{graphicx}
\usepackage{wrapfig}
\usepackage{pgf}
\usepackage{soul}
\usepackage{algorithm}
\usepackage{algpseudocode}
\usepackage{amsmath}
\usepackage{amsfonts}
\usepackage{mathtools}
\usepackage{amssymb}
\usepackage{mathrsfs}
\usepackage{cite}
\usepackage{subfigure}
\usepackage{tablefootnote}
\usepackage{bbm}
\usepackage{epstopdf}
\usepackage{threeparttable}
\everymath{\displaystyle}
\newtheorem{theorem}{Theorem}%[section] % This creates a theorem environment named "theorem"
\ifCLASSINFOpdf
\else
\fi
\begin{document}
%\title{Optimizing Alarm Scenario Performance through Spatially-based Transmission Thresholds for the IIoT Alarm Scenario}
%\title{Optimization of Transmission Thresholds for an IIoT Alarm Scenario}
\title{Configuring Transmission Thresholds in IIoT Alarm Scenarios for Energy-Efficient Event Reporting}

\author{David E. Ru\'{i}z-Guirola, \IEEEmembership{Graduate Student Member, IEEE}, Onel L. A. L\'{o}pez, \IEEEmembership{Senior Member, IEEE}, \\ 
and Samuel Montejo-S\'{a}nchez, \IEEEmembership{Senior Member, IEEE} 
\vspace{-2ex}%, John Doe, \IEEEmembership{Senior Member, IEEE} and , \IEEEmembership{Senior Member, IEEE}% <-this % stops a space
%\thanks{M. Shell was with the Department of Electrical and Computer Engineering, Georgia Institute of Technology, Atlanta, GA, 30332 USA e-mail: (see http://www.michaelshell.org/contact.html).}% <-this % stops a space
\thanks{David E. Ru\'{i}z-Guirola and Onel L. A. L\'{o}pez are with the Centre for Wireless Communications, University of Oulu, Finland. \{{David.RuizGuirola, Onel.AlcarazLopez\}.  
Samuel Montejo-S\'{a}nchez is with the {Instituto Universitario de Investigaci\'{o}n y Desarrollo Tecnol\'{o}gico, Universidad Tecnol\'{o}gica Metropolitana}, Santiago, Chile. \{smontejo@utem.cl\}.

This work has been partially supported %Chile by ANID FONDECYT Iniciaci\'on No. 11200659, %SCC-PIDi-UTEM 
%FONDEQUIP-EQM180180, and Collaborative Research Activities between PIDi/UTEM and FIE/UCLV, in Brazil by CNPq (402378/2021-0, 305021/2021-4), Print CAPES-UFSC ``Automation 4.0'', and RNP/MCTIC (Grant 01245.010604/2020-14) 6G Mobile Communications Systems, and in 
by the Finnish Foundation for Technology Promotion and the Research Council of Finland (former Academy of Finland) 6G Flagship Programme (Grant Number: 346208), the Finnish Foundation for Technology Promotion,  the European Commission through the Horizon Europe/JU SNS project Hexa-X-II (Grant Agreement no. 101095759), and in Chile, by ANID FONDECYT Regular No.1241977.  
%Manuscript received April 19, 2005; revised August 26, 2015.
}}}


\markboth{Journal of \LaTeX\ Class Files%,~Vol.~14, No.~8, August~2021
}%
{Shell \MakeLowercase{\textit{et al.}}: Bare Demo of IEEEtran.cls for IEEE Journals}



\maketitle
\vspace{-3ex}
\begin{abstract} 
Industrial Internet of Things (IIoT) applications involve real-time monitoring, detection, and data analysis. This is challenged by the intermittent activity of IIoT devices and their limited battery capacity. Indeed, the former issue makes resource scheduling and/or random access difficult, while the latter constrains IIoT devices' lifetime and efficient operation. 
In this paper, we address interconnected aspects of these issues. Specifically, we focus on extending the battery life of IIoT devices sensing events/alarms by minimizing the number of unnecessary transmissions. Note that when multiple devices access the channel simultaneously, there are collisions, potentially leading to retransmissions, thus reducing energy efficiency.
We propose a threshold-based transmission-decision policy based on the sensing quality  and the network spatial deployment. We optimize the transmission thresholds using several approaches such as successive convex approximation, block coordinate descent methods, Voronoi diagrams, explainable machine learning, and algorithms based on natural selection and social behavior. 
Besides, we propose a new approach that reformulates the optimization problem as a $Q$-learning solution to promote  adaptability to system dynamics. 
Through numerical evaluation, we demonstrate significant performance enhancements in complex IIoT environments, thus validating the practicality and effectiveness of the proposed solutions. 
We show that Q-learning performs the best, while the block coordinate descending method incurs the worst performance. Additionally, we compare the proposed methods with a benchmark assigning the same threshold to all the devices for transmission decision.  
Notably, in low-density scenarios, all the proposed methods outperform the benchmark. On the other hand, successive convex approximation, Voronoi-(i), the K-nearest neighbors-based, and Q-learning outperform the benchmark, while the remaining methods attain similar performance in high-density scenarios. 
Compared to the benchmark, up to 94\% and 60\% reduction in power consumption are achieved in low-density and high-density scenarios, respectively. 

\end{abstract}


% \fi
\IEEEpeerreviewmaketitle

\section{Introduction}

\IEEEPARstart{T}{he} Industrial Internet of Things (IIoT) %has a significant role in creating 
aims to %support a niche of 
connect 
Industry 4.0 %IoT applications that encompass 
and thus support 
predictive maintenance, asset tracking, smart manufacturing, energy management, environmental monitoring, health and safety monitoring, remote control, smart grid management, and agricultural automation~\cite{shi2020smart}. 
These applications enhance efficiency, safety, and sustainability across several industries, \textit{e.g.}, manufacturing, agriculture, energy, and logistics. 
Indeed, by incorporating intelligent attributes, including the utilization of real-time data and automation for process optimization and decision-making, IIoT networks allow cyber-physical systems to operate proactively and efficiently~\cite{chatzieleftheriou2022online}. %, providing intelligent services. 
%Among this services, factories with a smart approach to manufacturing align with the Industry 4.0 vision by promoting cooperative and effective manufacturing processes~\cite{chatzieleftheriou2022online}. 
%These applications enhance efficiency, safety, and sustainability across several industries, \textit{e.g.}, manufacturing, agriculture, energy, and logistics, by leveraging real-time data and automation to optimize processes and decision-making. 

%As %wireless devices %become more prevalent
%proliferate, 
IIoT networks are steadily evolving, driving massive data collection, analysis, and exploitation for control and sensing~\cite{Itimer}. In addition, the corresponding applications are often latency-sensitive and handle large amounts of IIoT entities sharing scarce communication resources, e.g., time and bandwidth, resulting in severe co-channel interference. Unfortunately, coordinating the transmissions to mitigate the interference incurs significant overheads, especially because the payload size in IIoT is usually small~\cite{alliance2018white}. Moreover, the sporadic activation of devices leads to highly inefficient coordination. Balancing resource scheduling without prior knowledge about when transmission resources are required and managing interference in large IIoT networks pose significant challenges~\cite{belmega2022online}. Additionally, due to the limited battery capacity of IIoT devices (IIoTDs), improving energy efficiency and incorporating energy harvesting sources are crucial~\cite{ul2022learning}. 

Note that the spatial arrangement of nodes in an IIoT network significantly impacts event detection accuracy and the effectiveness of alarm notifications. Indeed, one can enhance coverage, minimize blind spots, and optimize data transmission, thereby improving the overall efficiency of the network, by strategically deploying the IIoTDs~\cite{riihijarvi2010modeling}.  Also, one can leverage available information on the spatial correlation among devices to enhance transmission policies upon an event triggering. For instance, tools such as Voronoi diagrams can help identify specific coverage areas for individual IIoTDs within a given space, aiding in spatial analysis~\cite{lopez2023statistical}. Specifically, a clustering-based distributed learning solution for a medium access scheme is proposed in~\cite{ul2022learning} to tackle the problem where a set of sensors may  communicate a joint observation. 
The complexity lies in the limited signaling and shared information constraints, distinct from related research on exploiting sensor activity correlations or reliable alarm message transmission. In~\cite{raghuwanshi2023channel}, the authors propose a novel channel scheduling method by leveraging the spatial correlation between device activations. 
Integrating spatial correlation information and employing clustering-based techniques offer promising avenues to optimize transmission policies and enhance the system's energy efficiency and successful event detection.

%In the meantime, additional techniques that rely on 
Data-driven techniques are also suggested, \textit{e.g.}, in~\cite{intro1,intro2,intro3,intro4,intro5}. The authors in~\cite{intro1} propose a security framework that correlates activity across space and time to detect transmission patterns using data mining and supervised machine learning (ML), reaching 99\% accuracy. In~\cite{intro2}, the authors employ data capturing the correlations between devices to develop a random forest-based predictor for energy consumption in solar-powered nodes, leading to a 14\% reduction in prediction error.  %when using the %most relevant 
%training data based on the correlation between devices. 
The age-of-information (freshness of information) optimization problem is tackled using spatial correlations among IoT devices in~\cite{intro3}. %, while proposed policies outperform the state of the art. 
The authors in~\cite{intro4} present a framework for deriving high-level industrial events from low-level raw data streams, which can be used for online correlation analysis with high-level process events into a process event log. Meanwhile, in~\cite{intro5}, a statistical model is proposed for joint sensor identification and channel estimation using the least absolute shrinkage and selection operator joined to the 
orthogonal matching pursuit.   

%While some IIoT projects employ specific mathematical approaches, a fundamental problem remains in the utilization of useful methods for addressing communication issues from a general perspective~\cite{belmega2022online}. 
Despite the previous IIoT-related research projects, complex issues such as scalability, resource efficiency, and energy efficiency remain open challenges in implementing correlation-based strategies effectively~\cite{belmega2022online}. Addressing these issues is essential to fully optimize IoT connectivity performance as illustrated in this work. 
\begin{center}
    \begin{table}[t!]
    \begin{threeparttable}
    \caption{List of Symbols}
    \label{Symbol}
    \centering
    \begin{tabular}{ll}%p{6.8cm}}
    \hline            
    \textbf{Symbol} &  \textbf{Description} \\ %\midrule
    \hline   
    %$a \times b$   &Network area dimensions (rectangular)\\
    $d_{i,j}$   &Distance between $i^{\text{th}}$ event and an IIoTD $j$ at ($x_j,y_j$)\\ 
    $r_{j,h}$   &Distance between two IIoTDs at ($x_j,y_j$) and ($x_h,y_h$)\\
    %$i^{th}$ and $j^{th}$\\
    %$d_k$   &Distance from point $k$ to its Voronoi polygon\\
    %$d_x$   &Distance in x-axis\\
    %$d_y$   &Distance in y-axis\\
    $\mathbb{E}(\cdot)$   &Expected value operator\\
    %$f_X()$, $f_Y()$   &Distance probability function in x-axis and y-axis\\ 
    %&dimensions, respectively\\
    %$F_Z()$   &CDF of random variable Z\\
    $f_X(\cdot)$, $F_X(\cdot)$   &PDF$^*$ and CDF$^*$ of random variable $X$\\ 
    %&dimensions, respectively\\
    %$F_Z()$   &CDF of random variable Z\\
    %PDF and CDF of random variable X
    $g$   &Maximum number of iterations for the algorithms\\
    $\mathcal{J}$ &Set of IIoTDs\\
    $M$   &Number of clusters in k-nearest neighbors\\
    P   &Transition probability from inactive to active
state\\
    $p(\cdot)$   &Sensing power function\\
    $P_{\text{col}}$   &Collision probability\\
    $P_{\text{e}}$   &Error probability\\
    $P_{\text{miss}}$   &Miss-detection probability\\
    $\Pr\left(A_j\right)$   &Steady-state probability of IIoTD $j^{th}$ being active\\
    $\Pr\left(A_j|A_h\right)$   &Probability of IIoTD $j^{th}$ active if IIoTD $h^{th}$ is active\\
    %$q$         &Burst factor for consecutive data packets\\
    $Q(\cdot)$   &Lookup $Q$-table\\
    %$Q^*(s, a)$     &Optimal $Q$-function\\
    %$s$   &System state  $s\in S$\\
    $S$   &System space\\
    $(s,a)$     &State-action pair\\
    $\mathcal{T}(\cdot)$   &First-order Taylor serie approximation operator %around $\delta_0$
    \\
    %$u$   &$(\text{max}(x_j;a-x_j))^2$\\
    %$v$   &$(\text{max}(y_j;b-y_j))^2$\\
    %$w$   &$u+v$\\
    $W(\cdot)$   &Expected power consumption per IIoTD\\
    %$z$   &$d_{i,j}^2$\\
    $\alpha$    &Probability of event occurrence\\
    $\beta$   &Cardinality of population vector in GA$^*$\\
    $\xi$   &Network area\\
    $\eta$   &Control factor sensitivity for a given distance\\
    $\boldsymbol{\delta}$   &IIoTD transmission threshold vector\\
    $\nabla$   &First-order derivative operator\\ 
    $\mu_1,\mu_2$   &Balance utility/cost factors\\
    $\gamma_j$   &Energy efficiency factor of the $j^{th}$ IIoTD\\
    $\rho_j$   &Collision effect factor of the $j^{th}$ IIoTD\\
    $\sigma$   &Miss-detection factor\\
    %$\pi^*$   &Optimal policy\\
    $U$   &Cumulative discounted reward\\ %at  state $\tau$\\
    $\zeta^\tau$   &Discount factor at state $\tau$\\
    $r_{\tau+1}$   &Reward at a next state $\tau+1$\\
    $\pi$   &Policy employed within RL\\
    $\omega_\tau$   &Learning rate at state $\tau$\\
    $\epsilon$   &Non-stuck factor (random action factor)\\
    $\mathcal{O}(\cdot)$ &Big $\mathcal{O}$ notation\\
    $\Theta$     &Population size for GA$^*$ and PSO$^*$\\
    $\Omega$    &Voronoi distance\\
    \hline
\end{tabular}
\begin{tablenotes}
\item[*] {{Probability distribution
function (PDF), cumulative distribution
function (CDF), genetic algorithm (GA), and particle swarm optimization (PSO)}}.
\end{tablenotes}
\end{threeparttable}
\label{Symbol}
\end{table}
\end{center}
\vspace{-2.5ex}

\vspace{-0.5ex}
In this paper, we %address energy efficiency associated with IIoT networks, the 
focus on efficiently allocating transmit resources in an IIoT scenario to prolong the battery life of the devices. 
The trade-off involving resource allocation and energy efficiency is a complex design problem that we tackle from a perspective that considers spatial and temporal correlations between devices. 
%To address these issues, this article discusses solutions to energy optimization problem from correlation perspective between for optimizing any network environment that is unknown or complex, using convex optimization techniques that have been extensively used to obtain computationally attractive (exact or approximate) solutions to originally difficult design problems. 
For this, %this paper explores solutions for optimizing energy consumption in IIoT network environments. 
we leverage convex optimization techniques and derive computationally efficient solutions. 
Specifically, the contributions of this work are summarized as follows: 
\begin{itemize}
    \item We propose that the IIoTDs adopt threshold-based transmission decisions based on the sensing quality and network spatial deployment. We assess the impact of such an approach on coverage area, energy efficiency, and successful event detection.     
    %\item We determine the  impact of leveraging transmission threshold on network performance and reliability.     
    \item We propose %optimizing the activation threshold of devices for alarm event transmission relying on spatial correlation and event-triggering probability. Our approach involves 
    successive convex approximation (SCA), block coordinate descending (BCD), and different heuristic and data-driven solutions to optimize the transmission thresholds. The latter solutions include Voronoi diagrams, explainable machine learning, and algorithms based on natural selection and social behavior. 
    %ormulate the optimization problem and solve this optimization problem using a convex approximation to enhance network performance.
    %We optimize the activation threshold of devices for alarm event transmission based on spatial correlation and event triggering probability. 
    %\item We formulate the optimization problem and propose a convex approximation that is solved via successive convex approximation (SCA). 
    %\item We propose different heuristic and data-driven solutions to the optimization problem using Voronoi diagrams, explainable ML, and  algorithms based on natural selection and social behavior.  
    \item We reformulate the optimization problem as a reinforcement learning (RL) challenge and propose a $Q$-learning solution due to its ability to handle complex and dynamic environments where traditional optimization techniques struggle, and where finding an exact solution is computationally expensive or infeasible.  
    %that leverages the power of deep neural networks to enable more efficient and adaptive decision-making in complex scenarios where exploration is crucial to discover unknown optimal policies.  
    \item We improve energy efficiency and scalability in complex environments by exploiting the different approaches proposed in this paper. We show that RL performs the best, while BCD performs the worst. Moreover, all the proposed methods outperform the benchmark with equal transmission threshold setup for low-density scenarios, while they perform at least similarly to the benchmark for high-density scenarios. 
    %SCA, Voronoi-(i), K-nearest neighbors (KNN)-based, and RL outperform the benchmark for high density scenarios, while the remaining methods exhibit similar behavior. 
    Notably, energy consumption is reduced by up to 94\% for low-density scenarios and up to 60\% for high-density scenarios. 
    %The proposed methods outperform the benchmark for low density scenarios while for high density scenarios only SCA, Voronoi-(i), KNN-based and RL outperform the benchmark and the other have a similar behavior.      
\end{itemize}

The rest of the paper is organized as follows. Section II presents the system model and event influence analysis. Section III formulates the optimization problem. The proposed convex optimization solutions are presented in Section IV. In contrast,  solutions based on heuristic and RL are proposed in Section V.   
Section VI discusses the computational complexity of the proposed methods, while Section VII evaluates the proposed method through numerical simulations. Finally, we conclude the paper and list potential future work in Section VIII. Table~\ref{Symbol} lists the symbols used throughout this paper.

\section{System Model}

\begin{figure}[t!]
	\centering
	\includegraphics[width=0.95\columnwidth]{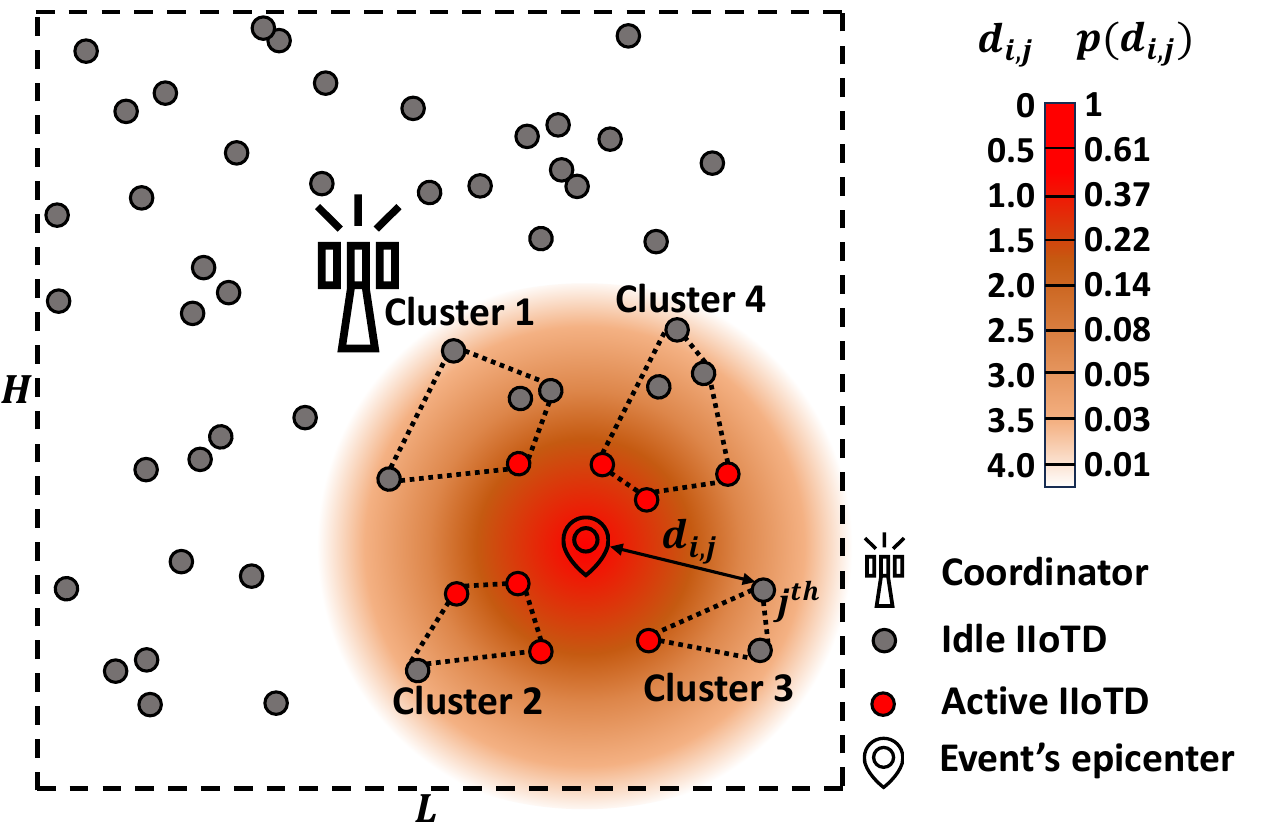}
	\vspace{-2mm}
	\caption{Illustration of an IIoT network where a coordinator controls and collects information from $N=55$ IIoTDs. The influence of an event on the surrounding IIoTDs is modeled by a probability activation function that decays with the distance from the event epicenter to the IIoTDs. %A two-state Markov model characterizes the IIoTD activation.
 }\vspace{-2mm}
	\label{figure1}
\end{figure}

We consider a rectangular coverage area, $\xi$, of dimensions $L\times H$, where a single coordinator/base station serves as the gateway for a set $\mathcal{J}$ %$N$ 
of $N$ short-range %machine-type communication devices (IIoTDs) 
IIoTDs %($j \in \mathcal{J}$)
as depicted in Fig.~\ref{figure1}. 
%Each device is awaiting for a triggering event, \textit{e.g.,} a moving object in motion detection applications or a fire in a fire-alarm system. While the IIoTD remains inactive (no alarm event activates it) the IIoTD holds in idle state ($I$), otherwise it switches to  
%State idle ($I$) represents the state in which the IIoTD remains inactive (no alarm event activates it) and 
%the active state ($A$) representing the activation by an alarm event and the need to send information to the coordinator/BS (red dots). 
These IIoTDs aim to detect the triggering of events, such as a moving object in motion detection applications or a fire in a fire-alarm system. %When the IIoTD remains inactive (no alarm event is triggered), the device stays in an idle state ($I$). Conversely, it transitions to the active state ($A$) when an alarm event activates it, indicating the need to transmit information to the coordinator/BS. %(depicted by red dots). 
The device switches from idle ($I$) to active ($A$) when an alarm event is detected, signifying the need for data transmission to the coordinator. 
The active devices send packets to the coordinator, which controls all the information exchange within its cell~\cite{FWuS}. For simplicity, we assume that all IIoTDs transmit with equal power. 

Assume that time is slotted in transmission time intervals (TTIs). 
%In a state $A$ (active), 
The {device}s report their sensing information in state $A$ (active), %with rate $R$, 
while no traffic is generated in state $I$ (idle). 
%To model the position of IIoTDs %we use Poisson point processes (PPPs) as typical nodes and events can be reasonably assumed to be stochastically deployed in the Euclidean plane. %~\cite{IoT_FWuS}. 
%Specifically, the IIoTDs are deployed according to a 2D homogeneous PPP $\Phi_{M}$ with density $\lambda_{M}$. On the other hand, the event epicenters are represented by a 2D homogeneous PPP $\Phi_{E}$ with density $\lambda_{E}$. The processes $\Phi_{M}$ and $\Phi_{E}$ are assumed to be independent, 
We assume that the coordinator knows each IIoTD location and that each event occurs uniformly in the area with probability ($\alpha$) in every TTI~\cite{thomsen2017traffic}. 
%Once the event occurs, the activation of each IIoTD is modeled as a two-state Markov chain, as illustrated in Fig.~\ref{figure1}. 
%%State $I$ represents the state in which the IIoTD remains inactive (no alarm event activates it) and the state $A$ represents the activation by an alarm event and the need to send information to the coordinator/BS. 
%The burst parameter $q \in [0,1]$ models the cases where data packets might be transmitted consecutively after the detection of an event. %For simplicity, $q$ may be set to zero.  
%Therefore, the steady-state probability of the $j^{th}$ device being in active state is %obtained as
%%\begin{equation}
    %\begin{array}{ll}
        %\Pr\left(A_j\right) = \frac{\text{P}_j}{\text{P}_j + (1- q_j)},
    %\end{array}
    %\label{1}
%\end{equation}
%represented by $\Pr\left(A_j\right)$, as the transition probability from $I$ to $A$~\cite{thomsen2017traffic}. %= {\text{P}_j},$
%where P$_j$ is the transition probability from $I$ to $A$~\cite{thomsen2017traffic}. 

%\begin{figure}[t!]
	%\centering
	%\includegraphics[width=0.6\columnwidth]{Model_traffic.pdf}
		%\vspace{-3mm}
	%\caption{Two-state Markov model for IIoTD activation.}
	%\label{figure2}
%\end{figure}

%\subsection{Influence of an Event Epicenter}\label{sec22}

%To capture the effect of a given event on a sensing IIoTD, we 
Let us define $p(d_{i,j})$ as a sensing power function that represents the impact of the $i^\text{th}$ event, triggered with epicenter $(x_{i},y_{i})$, on the $j^\text{th}$ IIoTD within the network area, in the two-dimensional Euclidean plane $\Re^{2}$. Here, $d_{i,j}$ indicates the distance separating them. %~\cite{thomsen2017traffic}. 
Moreover, $p(d_{i,j}) \rightarrow [0,1]$ %[0, \infty) \rightarrow [0,1]$
is non-increasing to mimic a decaying influence of events as the distance $d_{i,j}$ increases. 
As an example, Fig.~\ref{figure1} depicts 
the influence of an event epicenter on the surrounding IIoTDs. 
Potential functions to be used include exponential~\cite{thomsen2017traffic,FWuS},  linear~\cite{alves2021wireless}, piece-wise linear~\cite{hejselbaek2018empirical,yang2020resource}, and power-law decay~\cite{alves2021wireless} for general scenarios, while step~\cite{sun2019modeling} or sigmoidal~\cite{alves2021wireless} functions may be used for more stringent scenarios like indoor setups. In this paper, we use an exponentially decreasing function, \textit{i.e.,} $p(d_{i,j}) = e^{-\eta d_{i,j}}$~\cite{FWuS}. 
Here, the parameter $\eta$ ($\eta > 0$) controls the average sensitivity for a given distance. 
% ($\eta \in(0;1]$)
%Here, the different values of $\eta$ should be taking into account when the network is deployed. In fact, $\eta$ impacts directly on the network coverage, so for a given netwrok area, the value of $\eta$ and the device density are essential to reach the better and optimal possible coverage. Fig.~\ref{heatmap} shows the impact of different values of $\eta$ and device's density. Notice that as we increase the number of devices, maintaining the same $\eta$ the coverage area increases. Meanwhile, a smaller $\eta$ value helps for a more granulated coverage but smaller, while for a greater $\eta$ value the coverage area increases. However, as the coverage area increases the overlapping in coverage areas causes a challenge in network design. Thus, finding an optimal combination between devices coverage area looks a promising direction to maximize the coverage area while increasing energy efficiency in the network, i.e., decreasing overlapping. 

\begin{figure}[t!]
	\centering
	\includegraphics[width=\columnwidth]{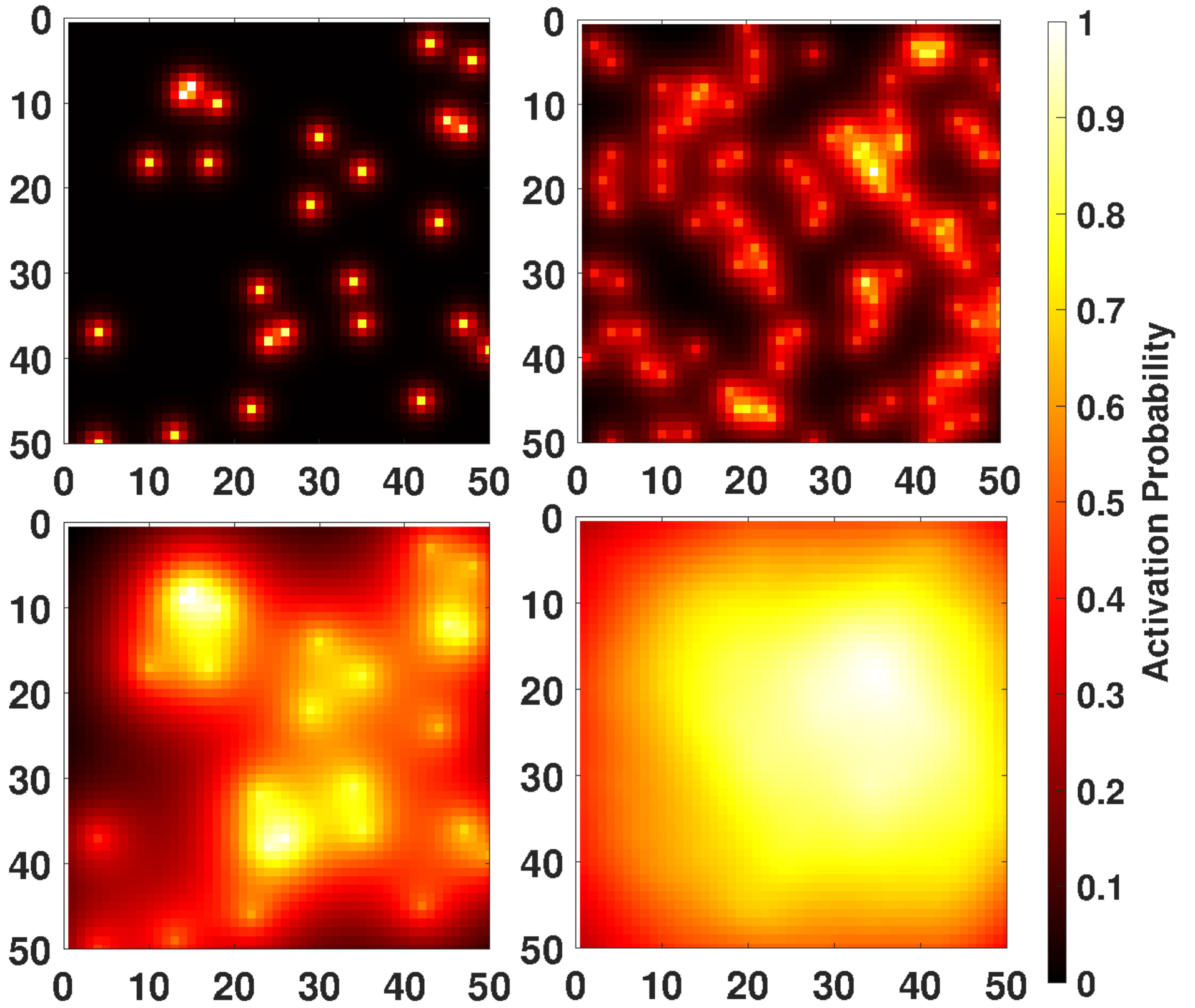}
		\vspace{-3mm}
	\caption{The coverage regions of an illustrative network deployment for a) (top left) $\eta = 1$ and 25 IIoTDs, b) (top right) $\eta = 1$ and 100 IIoTDs, c) (bottom left) $\eta = 0.1$ and 25 IIoTDs, and d) (bottom right) $\eta = 0.1$ and 100 IIoTDs. The activation probability shows the event detection coverage.}\vspace{-2mm}
	\label{heatmap}
\end{figure}
%\vspace{-3mm}

Note that $\eta$ directly influences network coverage, as depicted in Fig.~\ref{heatmap}. 
%Indeed, increasing the number of devices while maintaining $\eta$ constant leads to a larger coverage area. Conversely, 
A high $\eta$ value results in more granulated but smaller coverage, while a smaller $\eta$ value expands the coverage area. 
Note that the probability of detecting the event is low for the configuration depicted in Fig.~\ref{heatmap}(a) since there is a large uncovered area. On the other hand, the coverage area in Fig.~\ref{heatmap}(d) is the most extensive, while in Fig.~\ref{heatmap}(b-c) the coverage is medium.  
However, increasing coverage areas comes together with increasing  overlapping regions, which is not desired as it leads to increased collision probability and power consumption, contributing to the IIoTDs' battery depletion. %A bigger coverage area increases the activation probability but also the power consumption in the network and the collision probability. While a smaller coverage area increases the energy efficiency in the network but increases the miss-detection probability. 
Increasing (decreasing) the activation probability increases  (decreases) the power consumption of the IoTDs but also increases the collision (miss-detection) probability in the network. 
Thus, designing the IIoTDs' coverage areas for minimum overlap but realizing desired event detection capabilities extends overall coverage and enhances network energy efficiency. 
In this article, we explore a solution based on IIoTD clustering according to their spatial position/proximity, as depicted in Fig.~\ref{figure1}, thus optimizing resource management and enabling localized data processing. In addition, this aids in fault isolation and provides scalability to the IIoT system while improving energy efficiency.
Clustering IIoTDs enables the system to quickly detect and isolate faults within a particular cluster, reducing the impact on the overall network. Furthermore, it enhances scalability by managing clusters independently, simplifying the process of adding or removing devices without disrupting the entire system.

%Clustering IIoTDs allows the system to quickly identify and isolate faults within a specific cluster, minimizing the impact on the overall network. Furthermore, it enhances scalability by managing clusters independently, making it easier to add or remove devices without affecting the entire system. 

%\begin{figure}[t!]
	%\centering
	%\includegraphics[width=\columnwidth]{heatmap6.pdf}
		%\vspace{-3mm}
	%\caption{The coverage regions of one illustrative network deployment for a) (top left) $\eta = 1$ and 25 IIoTD, b) (top right) $\eta = 1$ and 100 IIoTD, c) (bottom left) $\eta = 0.1$ and 25 IIoTD, and d) (bottom right) $\eta = 0.1$ and 100 IIoTD. The activation probability shows the event detection coverage.}
	%\label{heatmap}
%\end{figure}

\section{Optimization Problem}

In the considered IIoT setup, devices make observations or measurements, which can be used to estimate the probability of successful signal reception, defined here as $p(d_{i,j})$. This probability is typically inferred from historical data or real-time measurements, considering factors such as signal strength and the deployment of IIoTDs.
Leveraging sensing signal probabilities, $p(d_{i,j})$, as a guide for transmission decisions holds promise for optimizing network efficiency. This accounts for the likelihood of successful signal reception at various distances, considering interference and IIoTD deployment. 

We propose setting (and optimizing) a threshold for each device %($\delta_j$) %based on the detection probability $p(d_{i,j})$. Devices can use this threshold %until conditions improve (\textit{e.g.}, reduced interference). This approach ensures that data transmission occurs when the likelihood of successful reception is high, 
to decide when to transmit data. We denote  
${\delta_j}$ as the transmission threshold that is dynamically chosen for the $j^\text{th}$ IIoTD. Then, the transmission probability for this $j^\text{th}$ IIoTD is given by 
\begin{equation}
\Pr\left(A_j\right) = \alpha \Pr\left( p(d_{i,j}) \geq {\delta_j}\right).
\end{equation}\label{P(Aj)}
\hspace{-1.5mm}Specifically, if the estimated $p(d_{i,j})$ for a given distance is lower than the preset threshold, the device avoids transmission. This is done with the hope that another device has a better observation of the event, thereby avoiding collisions in high-density scenarios, reducing re-transmissions, and conserving energy.  
%By setting transmission thresholds ($\delta$) based on $p(d_{i,j})$, we can enhance reliability and resource utilization, especially in energy-constrained scenarios like IoT. 
This aligns transmission strategies with real-world signal dynamics, promoting more responsive and energy-efficient communication. 

%Then, from \eqref{1} we have
%\begin{equation}
    %\begin{array}{ll}
        %\Pr\left(A_j\right) = \frac{\Pr\left( p(d_{i,j}) \geq {\delta_j}\right)}{1+\Pr\left( p(d_{i,j}) \geq {\delta_j}\right) - q}.
    %\end{array}
%\end{equation}
Herein, we aim %to formulate an optimization problem 
to minimize the energy consumption in the network by optimizing %$\delta_j$ for each IIoTD. 
%However, integrating considerations of energy constraints for IIoTDs add complexity, demanding communication strategies that effectively manage information exchange while optimizing energy efficiency.
%, without exceeding %a constraint values for 
%$\mathbb{E}(P_{\text{col}})$ and $\mathbb{E}(P_{\text{miss}})$ constraints, 
%an error constraint,
%and thus guarantee a quality of service (QoS). %In this sense, the correlation between IIoTDs in alarm scenarios plays an important role when optimizing the channel access approach. 
%Let us define 
\textbf{$\boldsymbol{\delta}$}$=[\delta_1, \delta_2, \dots \delta_N]^\mathbf{T}$. 
Specifically, 
the optimization problem is stated as \begin{subequations}
\begin{alignat}{4}
        \text{P}1: \text{ } &\min_{{\boldsymbol{\delta}}} \text{ } && W(\boldsymbol{\delta}) \\
        &\text{ } \text{s.t.} &&\mathbb{E}(P_{\text{e}}(\boldsymbol{\delta})) &&\leq \text{  } E,
        %& &&\mathbb{E}(P_{\text{miss}}) &&\leq \text{  } M,
\end{alignat}
\label{P1}
\end{subequations}
\hspace{-1.5mm}{where $P_{\text{e}}(\boldsymbol{\delta})$ is the error probability given the IIoTDs location and transmission threshold, $\mathbb{E}(P_{\text{e}})$ is its expected value with respect to the events' epicenters, and $E$ is the imposed error constraint.    
%without exceeding %a constraint values for 
%$\mathbb{E}(P_{\text{col}})$ and $\mathbb{E}(P_{\text{miss}})$ constraints, 
%an error constraint,
%to guarantee a quality of service (QoS), 
Meanwhile,  $W(\boldsymbol{\delta})$ depicts the expected power consumption per IIoTD in the network, which obeys} 
\begin{equation}
    \begin{array}{ll}
        W({\boldsymbol{\delta}}) %\propto 
        = \frac{1}{N}\sum_{j=1}^{N} \Pr\left(A_j\right). %\frac{\sum_{j=1}^{n}T_{\text{on}_{j}}}{n\times T_{\text{total}}}, 
    \end{array}
    \label{Eq3}
\end{equation}
%$\mathbb{E}(W) = \frac{\sum_{i=1}^{n}T_{\text{on}_{i}}}{n\times T_{\text{total}}}$, 
%being $n$ the number of IIoTDs in the network. 
%ON time $(T_{\text{on}})$ is equal to 
%\begin{equation*}
    %\begin{array}{ll}
        %T_{\text{on}} = \sum_{k=1}^{T_{\text{total}}} \mathcalm{1}_{\{p{(d)_{j}} \text{ } \geq \text{ } p_{\delta}\}},
    %\end{array}
%\end{equation*} 
%Here,  %the TTI is in the range $[0.125 \text{ } 0.25 \text{ } 0.5 \text{ } 1]$ ms and 
%$p_{\delta}$ is the  sensing threshold chosen for each IIoTD (optimization variable).
Here, we adopt a simplified energetic model, which is calculated as the proportion of time that the devices are in an active state. We disregard power consumption during idle periods due to its negligible impact on the overall analysis since, in IIoTDs, the primary battery depletion factor comes from radio interface usage.     
Notice that we assume one transmission window (Tx) per TTI, so more than one IIoTD trying to access the medium at the TTI slot time ($k^\text{th}$) is translated into collision. 

Collisions occur when multiple devices transmit their shared observations simultaneously, leading to data loss and increasing the expected error probability in our scenario. 
However, note that failing to detect the presence of an event also increases the  error probability. 
Given an IIoTD deployment, the event error probability $P_{\text{e}}$  
%(x_i,y_i)$ 
%(i^\text{th})$
captures collision ($P_{\text{col}}$) and miss-detection ($P_{\text{miss}}$) probabilities after the occurrence of an event, \textit{i.e.}, 
\begin{equation}
    \begin{array}{rl}
        %P_{\text{e}}(i^\text{th}) = P_{\text{col}}(i^\text{th}) + P_{\text{miss}}(i^\text{th}) \vspace{1mm}
         P_{\text{e}}(\boldsymbol{\delta}) = P_{\text{col}}(\boldsymbol{\delta}) + P_{\text{miss}}(\boldsymbol{\delta}), \vspace{1mm}
    \end{array}
\end{equation} 
since %the outcome is the same, \textit{i.e.,} the device is unable to send an event detection message. 
%We assume that an error has occurred in the network when an event occurs in a specific TTI, and 
the coordinator does not receive information about this event in both cases.  
The miss-detection probability %($%P_{\text{miss}} P_{\text{miss}}(x_i,y_i)$) 
is calculated as 
\begin{equation}\label{Pmiss}
    \begin{array}{rl}
        %P_{(\text{miss}|(x_i,y_i))} &= 
        %P(p{(d_{i,1})} < p_{\delta} \cup p{(d_{i,2})} < p_{\delta} \cup \dots \cup p{(d_{i,n})} < p_{\delta}),\\
        %= 
        P_{\text{miss}}(\boldsymbol{\delta}) =  \alpha \prod_{j=1}^{N} \Pr(p{(d_{i,j})} < {\delta_j})
        %\mathbb{E}(P_{\text{miss}}) &\propto \alpha P_{(\text{miss}|(x_i,y_i))}
    \end{array}
\end{equation}
due to the event not being detected by any sensor (miss-detection). Conversely, the  probability of message notification losses due to collisions when several, more than one IIoTD, transmit at the same time %(more than one IIoTD tries to send information at the same time) %($P_{\text{col}}$) %given an event with epicenter in position $(x_i,y_i)$ is calculated as 
is given by  
\begin{align}\label{Pcol}
        P_{\text{col}}(\boldsymbol{\delta}) = \alpha - P_{\text{miss}}(\boldsymbol{\delta}) - P_{\text{suc}}(\boldsymbol{\delta}),
        %P_{\text{col}} = 1 - \prod_{j=1}^{N} \Pr\left(p{(d_{i,j})} < {\delta_j}\right) + \dots \nonumber\\
        %- \sum_{h=1}^{N}\left(\Pr\left(p{(d_{i,h})} \geq {\delta_h}\right) \prod_{j=1,j\neq h}^{N} \Pr\left(p{(d_{i,j})} < {\delta_j}\right)\right),
        %\mathbb{E}(P_{\text{col}}) &\propto \alpha P_{(\text{col}|(x_i,y_i))}
\end{align} 
%Hence, the expected collision probability is 
%\begin{equation}
%    \begin{array}{ll}
%        \mathbb{E}(P_{\text{col}}) \propto \alpha \int_0^{a}\int_0^{b} \frac{1}{ab} P_{\text{col}}(x_i,y_i) \partial x \partial y.
%    \end{array}
%\end{equation} 
%$\mathbb{E}(P_{\text{col}}) \propto \alpha P_{(\text{col}|(x_i,y_i))}$. 
%Hence, %$ P_{\text{e}}(x_i,y_i) = P_{\text{col}}(x_i,y_i) + P_{\text{miss}}(x_i,y_i)$ is given by 
%\begin{equation}
    %\begin{array}{cl}
        %P_{\text{e}}(x_i,y_i) = P_{\text{col}}(x_i,y_i) + P_{\text{miss}}(x_i,y_i) \vspace{1mm} \\
        %= 1 - \sum_{h=1}^{N}\left(\Pr\left(p{(d_{i,h})} \geq {\delta_h}\right) \prod_{j=1,j\neq h}^{N} \Pr\left(p{(d_{i,j})} < {\delta_j}\right)\right), 
        %\mathbb{E}(P_{\text{col}}) &\propto \alpha P_{(\text{col}|(x_i,y_i))}
    %\end{array}
%\end{equation} 
where $P_{\text{suc}}$ is the success probability. Therefore, we have that   
\begin{align}\label{Pe}
        P_{\text{e}}(\boldsymbol{\delta}) = \alpha - P_{\text{suc}}(\boldsymbol{\delta}).  
        %\mathbb{E}(P_{\text{col}}) &\propto \alpha P_{(\text{col}|(x_i,y_i))}
\end{align}
Herein, $P_{\text{suc}}$ refers to the union of the detection events from one and only one of the $N$ sensors described as 
\begin{align}\label{Psuc}
        P_{\text{suc}}(\boldsymbol{\delta}) = \alpha \sum_{h=1}^{N}\Big(\Pr\left(p{(d_{i,j})} \geq {\delta_h}\right) \prod_{j\neq h}^{N} \Pr\left(p{(d_{i,j})} < {\delta_j}\right)\Big). 
        %\mathbb{E}(P_{\text{col}}) &\propto \alpha P_{(\text{col}|(x_i,y_i))}
\end{align}
Then, the expected value in the %$a \times b$ 
network area %($\xi$) 
is given by  
\begin{equation}
    %\begin{array}{l}
        \mathbb{E}(P_{\text{e}}(\boldsymbol{\delta})) = %\mathbb{E}(P_{\text{col}}(\boldsymbol{\delta})) +\mathbb{E}(P_{\text{miss}}(\boldsymbol{\delta}))
        \alpha - \mathbb{E}(P_{\text{suc}}(\boldsymbol{\delta}))
        = %\hspace{2mm} 
        %\alpha %\int_0^{a}\int_0^b  
        \int_\xi %\oint_\xi
        \frac{1}{|\xi|} P_{\text{e}}(\boldsymbol{\delta}) \partial \xi.  %x \partial y.
    %\end{array}
    \label{expterror}
\end{equation}
%$\mathbb{E}(P_{\text{miss}}) \propto \alpha \int_0^{a}\int_0^b \frac{1}{ab} P_{(\text{miss}|(x_i,y_i))} \partial x \partial y$.
%Collision and miss-detection probabilities are the main challenges in this scenario. This pertains to situations where multiple devices attempt to transmit their shared observation simultaneously, resulting in signal interference. Additionally, the likelihood of some devices failing to detect the shared message's presence further complicates the reliable delivery of this collective information. 
%Collision and miss-detection probabilities 
%Error probability constitute the central challenge in this scenario. %, particularly when viewed in the context of energy consumption.   

\section{Convex Optimization-based solution}

%We have that

%\begin{equation}
    %\begin{array}{lll}
        %\min_{p_{\delta}} &W(p_{\delta_j}) \\
        %\text{ } \text{s.t.} &\mathbb{E}(P_{\text{col}}) &\leq \text{  } C\\
        %&\mathbb{E}(P_{\text{miss}}) &\leq \text{  } M\\
    %\end{array}
%\end{equation}

%where $C$ and $M$ are the collision and miss-detection constraints, respectively, 

%While assuming a different threshold for every IIoTD and $p(d) = e^{-d}$ 
Substituting $\alpha \Pr \textstyle \left( e^{-\eta d_{i,j}} > \delta_j\right)$
into \eqref{Eq3}, the expected power consumption in the network obeys 
\begin{equation}
    \begin{array}{ll}
        W(\boldsymbol{\delta}) %\propto 
        = \frac{\alpha}{N}\sum_{j=1}^{N} {\Pr\left( e^{-\eta d_{i,j}} > \delta_j\right)}.%{\Pr\left( e^{-\eta d_{i,j}} > \delta_j\right) + (1- q)}. 
        %= \frac{\alpha}{N}\sum_{j=1}^{N} \frac{\Pr\left( e^{-\eta d_{i,j}} > \delta_j\right)}{\Pr\left( e^{-\eta d_{i,j}} > \delta_j\right) + (1- q)}.
    \end{array}
\end{equation}
This is monotonically increasing %for values of 
on $\Pr\left( e^{-\eta d_{i,j}} > \delta_j\right)$,  
%with  
%assuming $\eta = 1$~\cite{FWuS},
%$\Pr \textstyle \left( e^{-\eta d_{i,j}} > \delta_j\right) \in [0,1]$,  
which in turn is exponentially decreasing on $\delta_j$. 
Then, the optimization problem can be reformulated as minimizing $\Pr \textstyle \left( e^{-\eta d_{i,j}} > \delta_j\right)$ equivalent to $\Pr\left( {d_{i,j}} < -\mathrm{ln}(\delta_j)/\eta \right)$. 
%Now, the expected value of the function $\Pr\left( {d_{i,j}} < -\mathrm{ln}(\delta_j)\right)$ depends on a random variable 
Note that 
$d_{i,j} = \textstyle \sqrt{(X-x_j)^2 + (Y-y_j)^2}$, where $X$ and $Y$ are random variables. Herein, without losing generality, we assume that $X$ and $Y$ are equally distributed in $[0,L]$ and $[0,H]$, respectively, and $(x_j,y_j)$ is the known position of the device $j$. % in a rectangular area $\xi = L \times H$. 

%From now on, we use $d_{i,j}$ and $d$, indistinctly, when referring to $d_{i,j}$. 
\begin{theorem}\label{cdf_theorem}
%In order to calculate the cumulative distribution function (CDF) of $d_{i,j}$, we first model the variable 
Let 
$Z = d_{i,j}^2 = (X-x_j)^2 + (Y-y_j)^2$, %for simplicity, while $u = (\text{max}(x_j;L-x_j))^2$, $v = (\text{max}(y_j;H-y_j))^2$,  and $w = u + v$. The CDF of $z$ has a closed-form and is calculated as (see proof in Appendix~\ref{append1})
then 
\begin{align}\label{cdf}
    F_Z(z) = \frac{2z}{\xi}\left( \text{arcsin}\left(\sqrt{\frac{u}{z}}\right) + \frac{1}{2}\sin{\left(2 \text{arcsin}\left(\sqrt{\frac{u}{z}}\right)\right)} \right), 
\end{align}%\vspace{-4ex}
where $u = \text{max}(x_j;L-x_j)^2.$ %+ \text{max}(y_j;H-y_j)^2.$
\end{theorem}
    
%\begin{proof}%[\textit{Theorem} \ref{cdf_theorem}]
    %The CDF of $z$ has a closed-form and is calculated (
\textit{Proof:} See proof in Appendix~A.
%\end{proof}

%The CDF of $z$ has a closed-form and is calculated as (see proof in Appendix~\ref{append1})
%\begin{align}\label{cdf}
    %F_z(z) = \frac{2z}{uv}\left( \text{arcsin}\left(\sqrt{\frac{v}{z}}\right) + \frac{1}{2}\sin{\left(2 \text{arcsin}\left(\sqrt{\frac{v}{z}}\right)\right)} \right).
%\end{align}
%However, the presence of the $\sin$ and $\arcsin$ functions in the expression makes the function in~\eqref{cdf} inherently non-convex and without a closed-form derivative. 

The function in~\eqref{cdf} is inherently non-convex   
%and without a closed-form derivative, 
and highly non-linear due to the significant 
oscillations and fluctuations introduced by the trigonometric functions, leading to multiple local minima and maxima. This makes \eqref{cdf} computationally expensive to analyze and optimize. 
Note however that %of~\eqref{cdf}
\begin{equation}\label{approximation_Fz}
    F_Z(z) \approx 1 - e^{-\frac{2z}{w}}, \text{ for }z \leq 200/\eta^2, 
\end{equation}
%, for $z \leq 200/\eta^2$, %the function $F_z(z)$ can be approximated as 
%is given by 
%$F_z(z) \approx 1 - e^{-\frac{2z}{w}}$ \cite{miller2012probability}
which is a practical sensing range for low-power IIoTD (less than 5 meters~\cite{ul2022learning}, corresponding to $\delta \in [10^{-4}, 1]$). 
\begin{figure*}[t!]\label{Fz}
	\centering
        \includegraphics[width=\columnwidth]{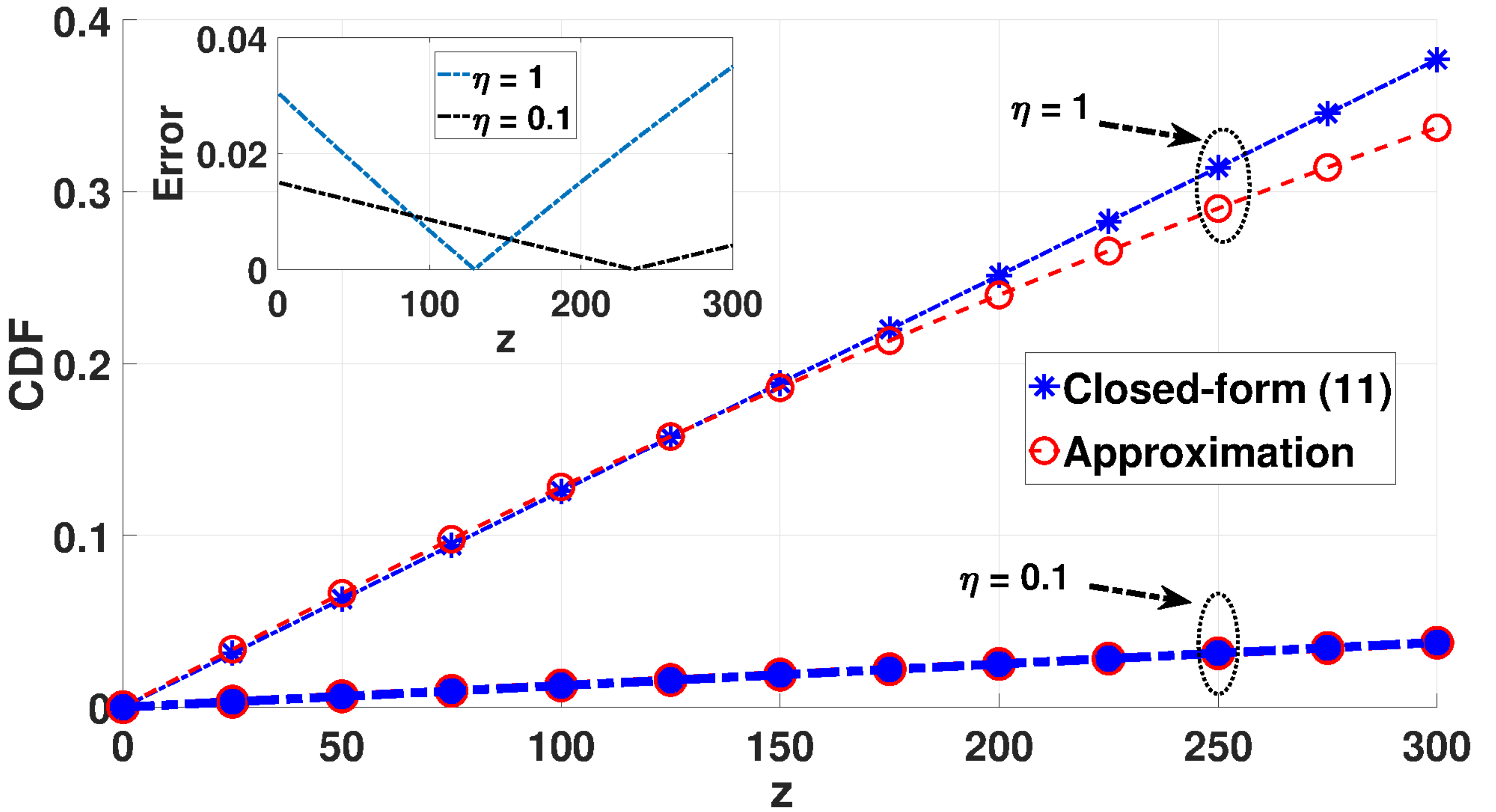}
        \includegraphics[width=\columnwidth]{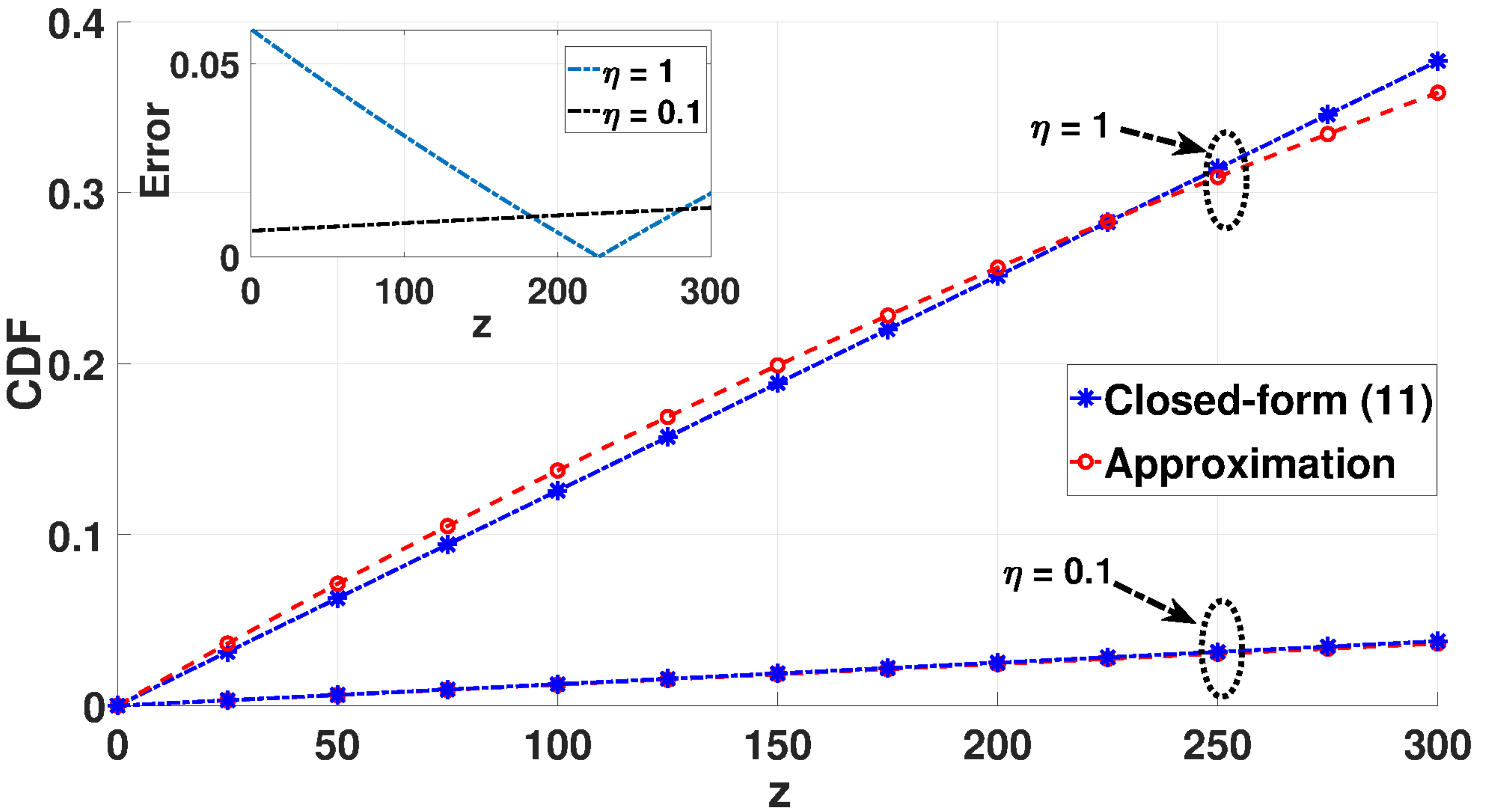}
        \vspace{-3mm}
	\caption{Approximation to \eqref{cdf} fo different $u,v$ and $\eta$ values, (left) $u,v = 25$ and (rigth)  $u,v = 50$.}\vspace{-2mm}
    \label{Fz}
\end{figure*}
Fig.~\ref{Fz} illustrates the accuracy of \eqref{approximation_Fz} for various values of $u$, $v$, and $\eta$, suggesting its suitability for reformulating the complexity of \eqref{P1} into a more manageable form. 
Therefore, $\Pr\left( {d_{i,j}} \leq -\mathrm{ln}(\delta_j)/\eta \right)$ can be reformulated as $\Pr\left( {z_{j}} \leq \mathrm{ln^2}(\delta_j)/\eta^2 \right) = F_z \left(\mathrm{ln^2}(\delta_j) /\eta^2\right)$, which is monotonically increasing, and P1 can be rewritten as
\begin{subequations}
\begin{alignat}{4}
        \text{P}2:\text{ }  &\min_{\boldsymbol{\delta}} \text{ } &&\sum_{j=1}^{N}{F_z \left(\mathrm{ln^2}(\delta_j)/\eta^2\right)} \\
        &\text{ } \text{s.t.} &&\mathbb{E}(P_{\text{e}}(\boldsymbol{\delta})) \leq \text{  } E.
        %& &&\mathbb{E}(P_{\text{miss}}) &&\leq \text{  } M,
    \end{alignat}
    \label{P2}
\end{subequations} 
\hspace{-1.5mm}Herein, $\mathbb{E}(P_{\text{e}}(\boldsymbol{\delta}))$ can be computed by substituting %\eqref{Pmiss}, \eqref{Pcol}, and 
\eqref{Psuc} into \eqref{expterror}, wherein 
%\begin{equation}
    %\begin{align}\label{EPmiss}
        %\mathbb{E}(P_{\text{miss}}(\boldsymbol{\delta})) %=& \prod_{j=1}^{N} (1 - P\left(d_{i,j} < \delta_j /\eta \right)) \nonumber\\
        %=& \prod_{j=1}^{N} \left(1 - F_z \left(\mathrm{ln^2}(\delta_j) /\eta^2 \right)\right) = \prod_{j=1}^{N} e^{-\frac{2\mathrm{ln^2}(\delta_j)}{w_j}} \nonumber\\
        %=& e^{-2 \sum_{j=1}^N \frac{\ln^2 \delta_j}{w_j}},
    %\end{align}
%and the expected collision probability: 
\begin{align}\label{EPsuc}
        \mathbb{E}(P_{\text{suc}}(\boldsymbol{\delta})) =& \alpha \sum_{h=1}^{N} \left(1 - e^{-\frac{2\mathrm{ln^2}(\delta_h)}{w_h}} \right) \prod_{j\neq h}^{N} e^{-\frac{2\mathrm{ln^2}(\delta_j)}{w_j}} \nonumber\\ 
        =& \alpha \sum_{h=1}^{N} \left(1 - e^{-\frac{2\mathrm{ln^2}(\delta_h)}{w_h}} \right) e^{-2 \sum_{j\neq h}^N \frac{\ln^2 \delta_j}{w_j}}\nonumber\\ 
        =& \alpha \sum_{h=1}^{N} e^{-2 \sum_{j\neq h}^N \frac{\ln^2 \delta_j}{w_j}} +  {\alpha}{N} e^{-2 \sum_{j=1}^N \frac{\ln^2 \delta_j}{w_j}}.  
%\end{align}
%\begin{align}\label{EPcol}
        %\mathbb{E}(P_{\text{col}}(\boldsymbol{\delta})) =& 1 - \mathbb{E}(P_{\text{miss}}) - \mathbb{E}(P_{\text{suc}}). 
\end{align} 
%\end{equation} 
Hence, 
\begin{align}\label{E_Pe_convex}
        \mathbb{E}(P_{\text{e}}(\boldsymbol{\delta})) %=& \alpha - \alpha \sum_{h=1}^{N} \left(1 - e^{-\frac{2\mathrm{ln^2}(\delta_h)}{w_h}} \right) \prod_{j\neq h}^{N} e^{-\frac{2\mathrm{ln^2}(\delta_j)}{w_j}} \nonumber\\ 
        %=& \alpha - \alpha \sum_{h=1}^{N} \left(1 - e^{-\frac{2\mathrm{ln^2}(\delta_h)}{w_h}} \right) e^{-2 \sum_{j\neq h}^N \frac{\ln^2 \delta_j}{w_j}} \nonumber\\ 
        =& \alpha - \alpha \sum_{h=1}^{N} e^{-2 \sum_{j\neq h}^N \frac{\ln^2 \delta_j}{w_j}} +  {\alpha}{N} e^{-2 \sum_{j=1}^N \frac{\ln^2 \delta_j}{w_j}}.
\end{align}
%Minimizing $ \textstyle \left( 1 - e^{-\frac{2z}{w}}\right)$ is equivalent to %maximizing the second term $e^{-\frac{2z}{w}}$, at the same time, maximizing a negative exponential is the same as 
%minimizing $z$ %the exponent $({2z}/{w})$. Now, 
%since 
%$w$ %is constant for a given scenario, since it 
%refers to the maximum distance between an event and a device and thus is constant for a given scenario. 
%Therefore, what we can optimize is the term th (the threshold for a device to be activated), \textit{i.e.}, the variable 
%Now, since 
%$z = \mathrm{ln^2}(\delta_j)$, 
%the problem can be formulated as 
%. Again, minimizing $z$, since $0 < $ th $< 1$ and we have a square logarithm, is equal to maximize (th). %as we can see in Fig.~\ref{active}. 
%Therefore the objective function can be changed as 
%\begin{figure}[t]
%	\centering
%	\includegraphics[width=\columnwidth]{active.pdf}
%		%\vspace{-3mm}
%	\caption{Mean threshold vs $\mathbb{E}(P_{\text{A}})$ normalized to $\alpha$.}
%	\label{active}
%\end{figure}
%\begin{subequations}
%\begin{alignat}{4}
%       \text{P}3: \text{ } &\max_{{\delta^T}} \text{ } &&{\delta_j}\\
%        &\text{ } \text{s.t.} &&\mathbb{E}(P_{\text{col}}) &&\leq \text{  } C,\\
%        & &&\mathbb{E}(P_{\text{miss}}) &&\leq \text{  } M,
%    \end{alignat}
%\end{subequations}
However, solving this optimization problem is still non-trivial since the objective and constraint functions are not convex (see proof in Appendix~B). 

Convex optimization techniques have been promising in addressing various challenges in wireless communication systems%. These approaches have been used in several optimization problems related to wireless communications, such as 
, including optimizing beamforming for enhanced signal transmission~\cite{convex1}, power consumption minimization~\cite{convex2}, and refining device localization methods for precise positioning~\cite{convex3}.
Moreover, interior point methods (IPM)~\cite{vargas1993tutorial,boyd2004convex}, which are known to converge in/with polynomial time/complexity, are commonly utilized for solving convex problems~\cite{ye2011interior,tondo2022optimal}. 
In the following subsections, we will discuss the fundamental principles of convex approximation methods used in this paper and how they address P2. 
Such an optimization problem can be solved by iterative approximation methods like SCA~\cite{boyd2020cvx,boyd2004convex} or BCD~\cite{gao2022low}, where an inner convex approximation in each iteration approximates the non-convex feasible set.  
By using the approximation methods, an approximate solution to the optimization problem in~\eqref{P2} can be attained using standard convex optimization tools 
such as CVX~\cite{boyd2020cvx} or fmincon of MATLAB~\cite{mathworksfmincon}. 
%These methods have a computational advantage for complex design problems, with a significant role in improving the performance and efficiency of wireless networks. 
 
\subsection{Successive Convex Approximation (SCA)}

\begin{algorithm}[t]
\caption{SCA}\label{alg_SCA}
\begin{algorithmic}[1]
\small % Change font size to \small
%\vspace{-2mm}
\State \textbf{Initialization:} Choose initial $\boldsymbol{\delta_0}$ 
\For{$k = 0, 1, 2, \ldots, g-1$}
    \State \textbf{Solve} the surrogate problem  using %the first-order Taylor series approximation 
    $\mathcal{T}(\cdot)$ around $\boldsymbol{\delta}_k$ such that  ${\boldsymbol{\hat{\delta}}}_k = \arg \min_{\boldsymbol{\delta}}  W(\boldsymbol{\delta}) \text{ } \text{s.t. } \mathbb{E}(P_{\text{e}}(\boldsymbol{\delta})) \leq E$
    %\State \hspace{5mm} ${\boldsymbol{\hat{\delta}}}^k = \arg \min$ P2
    \State \textbf{Update} with learning rate $\vartheta_k : \boldsymbol{\delta}_{k+1} = \boldsymbol{\delta}_{k} + \vartheta_k \left({\boldsymbol{\hat{\delta}}}_k-\boldsymbol{\delta}_k \right)$ 
    %\State \hspace{5mm} $\boldsymbol{\delta}^{k+1} = \boldsymbol{\delta}^{k} + \vartheta^k \left({\boldsymbol{\hat{\delta}}}^k-\boldsymbol{\delta}^k \right)$ 
    %\If{stopping criterion is satisfied}
        %\State \textbf{break} 
    %\EndIf
\EndFor
\State \textbf{Return} $\boldsymbol{\delta}_k$
\end{algorithmic}
\end{algorithm}

To solve P2 using SCA, we approximate the constraint function with its first-order Taylor series and iteratively optimize the objective function. Also, we use regularization to enforce convergence. Herein, we define $\mathcal{T}(\cdot)$ as the operator approximating an input function using its first-order Taylor series expansion.      
%The first-order Taylor series approximation for the constraint around $%\boldsymbol
%{\delta_0}$, which is a vector of initial values of $\boldsymbol{\delta}$ %$ =[{\delta}_1, {\delta}_2, \dots {\delta}_n]$ 
%that is updated in each SCA iteration by the resulting value obtained in the previous iteration, 
%with $\mathbb{E}(P_{\text{e}})$ evaluated in $\boldsymbol{\delta}$ for the error probability and $\nabla \mathbb{E}(P_{\text{e}}(%\boldsymbol
%{\delta_0}))$ for $\forall i \in \mathcal{J}$. 
The first-order Taylor series approximation is used to estimate the constraint around $\boldsymbol{\delta_0}$. This constraint is a vector of initial values of $\boldsymbol{\delta}$ that gets updated in each SCA iteration by the resulting value obtained in the previous iteration. The error probability and its first derivative is evaluated in $\boldsymbol{\delta}$, %and $\nabla \mathbb{E}(P_{\text{e}}(\boldsymbol{\delta_0}))$ 
%$\forall i \in \mathcal{J}$. 
Thus, it is given by
%\begin{equation}
    \begin{align}
        \mathcal{T}(\mathbb{E}(P_{\text{e}}(\boldsymbol{\delta}))) %= f(\delta_0) + (x-\delta_0)^\mathbf{T} \nabla f(\delta_0) \nonumber\\
            %&
            =& \mathbb{E}(P_{\text{e}}(\boldsymbol{\delta_0})) + (\boldsymbol{\delta}-\boldsymbol{\delta_0})^\mathbf{T} \nabla \mathbb{E}(P_{\text{e}}(\boldsymbol{\delta_0})), %\nonumber\\
            %=& \mathbb{E}(P_{\text{e}}(\boldsymbol{\delta_0})) - \boldsymbol{\delta_0}^\mathbf{T} \nabla \mathbb{E}(P_{\text{e}}(\boldsymbol{\delta_0})) + \boldsymbol{\delta}^\mathbf{T} \nabla \mathbb{E}(P_{\text{e}}(\boldsymbol{\delta_0})), 
    \end{align}
%\end{equation}
%where th$^T$ is the transposed vector of th values for each device $[\delta_1, \delta_2, \dots \delta_n]$. 
%Likewise, $f(\delta)$ would be $\mathbb{E}(P_{\text{e}})$ evaluated in $\delta$ for the error probability, and $\nabla f(\delta)$ for $\forall i \in N$ is 
where 
%\begin{align}
%\begin{equation}
    %\begin{array}{rl}
        %\nabla \mathbb{E}(P_{\text{miss}}(\boldsymbol{\delta_0})) =& \prod_{j=1}^{N} \frac{4e^{-\frac{2\mathrm{ln^2}({\delta}_j)}{w_j}}}{w_i \delta_i/ \mathrm{ln}(\delta_i)} = \frac{4e^{-2 \sum_{j=1}^N \frac{\ln^2 \delta_j}{w_j}}}{w_i \delta_i/ \mathrm{ln}(\delta_i)},\\
    %\end{array}
%\end{equation}
%likewise 
%\begin{align}
    %\begin{array}{ll}
        %\nabla \mathbb{E}(P_{\text{col}}(\boldsymbol{\delta_0})) =& \sum_{h=1}^{N} \nabla \mathbb{E}(P_{\text{miss}}(\boldsymbol{\delta_0})) - \sum_{h\neq i}^{N} \prod_{j\neq h}^{N} \frac{4e^{-\frac{2\mathrm{ln^2}({\delta}_j)}{w_j}}}{w_i \delta_i/ \mathrm{ln}(\delta_i)} \nonumber \\
        %=& \sum_{h=1}^{N} \nabla \mathbb{E}(P_{\text{miss}}(\boldsymbol{\delta_0})) - \sum_{h\neq i}^{N} \frac{4e^{-2 \sum_{j\neq h}^N \frac{\ln^2 \delta_j}{w_j}}}{w_i \delta_i/ \mathrm{ln}(\delta_i)}.
        %&+ \sum_{h=1}^{N} \prod_{j=1}^{N} \left(\frac{4e^{-\frac{2\mathrm{ln^2}({\delta}_j)}{w_j}} \mathrm{ln}(\delta_i)}{w_i \delta_i}\right),
%\end{align}
%Hence,
%Hence, by algebraic manipulation we arrive at
\begin{align}
    %\begin{array}{ll}
        %\nabla \mathbb{E}(P_{\text{e}}(\boldsymbol{\delta})) =& \sum_{h=1}^{N+1} \nabla \mathbb{E}(P_{\text{miss}}(\boldsymbol{\delta_0})) - \sum_{h\neq i}^{N} \prod_{j\neq h}^{N} \frac{4e^{-\frac{2\mathrm{ln^2}({\delta}_j)}{w_j}}}{w_i \delta_i/ \mathrm{ln}(\delta_i)} \nonumber\\ 
        %=& \sum_{h=1}^{N+1} \nabla \mathbb{E}(P_{\text{miss}}(\boldsymbol{\delta_0})) - \sum_{h\neq i}^{N} \frac{4e^{-2 \sum_{j\neq h}^N \frac{\ln^2 \delta_j}{w_j}}}{w_i \delta_i/ \mathrm{ln}(\delta_i)}. 
        \!\nabla_{\delta_j} \mathbb{E}(P_{\text{e}}(\boldsymbol{\delta}))\! &=\!  \frac{4\alpha \mathrm{ln}(\delta_j)}{w_j \delta_j} \!\left( \sum_{h=1}^{N-1}\! e^{-2 \sum_{i\neq h}^N\! \frac{\ln^2 \delta_i}{w_i}} \!\!-\! e^{-2 \sum_{i=1}^N\! \frac{\ln^2 \delta_i}{w_i}} \!\!\right).
\end{align}

Algorithm~\ref{alg_SCA} depicts the SCA algorithm for each iteration $k$. 
Note that the complexity when implementing SCA is $\mathcal{O}(g f({\boldsymbol{\delta}}))$, %^\mathbf{T})))$, with \textbf{$\mathbf{\delta}$}$^\mathbf{T}=[\delta_1, \delta_2, \dots \delta_N]^\mathbf{T}$, 
where $g$ is the maximum number of iterations.

\subsection{Block coordinate descending method (BCD)}

\begin{algorithm}[t]
\caption{BCD}\label{alg_BCD}
\begin{algorithmic}[1]
\small % Change font size to \small
%\vspace{-2mm}
\State \textbf{Initialization:} Choose initial $\boldsymbol{\delta_0}$ 
\For{$k = 0, 1, 2, \ldots, g-1$}
    \For{$j = 1, 2, \ldots, N$}
        \State \textbf{Solve} P2 for variable ${{\hat{\delta_j}}}$ with $\delta_{h\neq j}$ fixed, i.e., ${{{\hat{\delta_j}}} = \arg \min_{\delta_j}}\text{ } W(\boldsymbol{\delta})\text{ } \text{s.t. } \mathbb{E}(P_{\text{e}}(\boldsymbol{\delta})) \leq E$
        %\State \hspace{5mm} ${{\hat{\delta_j}}}^k = \arg \min$ P2 
    \EndFor
    %\State \textbf{Update}  
        %\State \hspace{5mm} $\boldsymbol{\delta}^{k+1} = {\boldsymbol{\hat{\delta}}}^k$
    \State $\boldsymbol{\delta}_{k+1} = {\boldsymbol{\hat{\delta}}}_k$
    %\If{stopping criterion is satisfied}
        %\State \textbf{break} 
    %\EndIf
\EndFor
\State \textbf{Return} $\boldsymbol{\delta_k}$
\end{algorithmic}
\end{algorithm}
%\vspace{-2mm}

%In addition, we use the block coordinate descending (BCD) method to solve the problem in~\eqref{P2}. 
%Particularly, i
In the conventional BCD framework, the formulated non-convex sparse recovery problem can be decomposed into small-scale sub-problems after exploiting the least absolute shrinkage and selection operator-based regularization~\cite{gao2022low}. Subsequently, the variables in each
sub-problem can be optimized %according to the closed-form solution 
sequentially with variables from other sub-problems kept fixed. 
Herein, we also approximate the constraint function with its first-order Taylor series, as in SCA, but in this case, the function has just one variable in each iteration, %$\forall j \in \mathcal{J}$, 
while the rest are fixed.  
The method runs $N$ iterations in which the problem is solved for one variable at each step, fixing the rest of $(N-1)$ variables, running $N$ sequential problems with one variable in an inner loop, and repeating the method $g$ times (maximum iteration). 
Algorithm~\ref{alg_BCD} summarizes BCD algorithm and note that its complexity is $\mathcal{O}(gN f({\boldsymbol{\delta}}))$.
%Therefore, the complexity when using BCD is $\mathcal{O}(gN f({{\delta}}))$.

\begin{figure*}[t!]
	\centering
	\includegraphics[width=0.65\columnwidth]{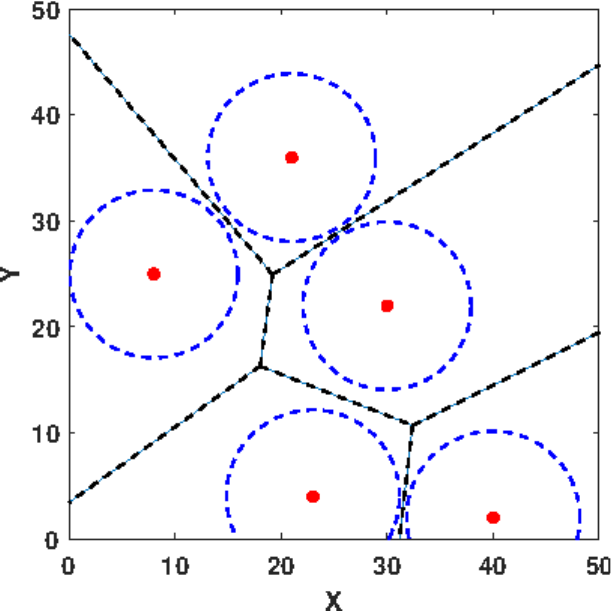}
        \includegraphics[width=0.65\columnwidth]{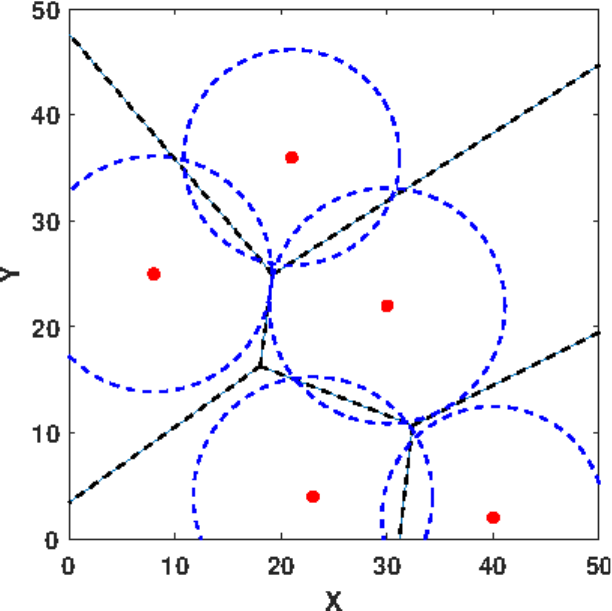}
        \includegraphics[width=0.65\columnwidth]{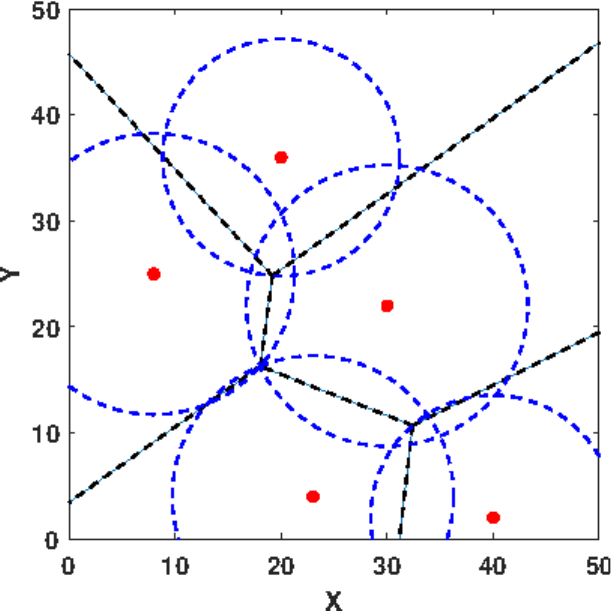}
        \vspace{-2mm}
	\caption{Illustration of a Voronoi diagram for a grid IIoT deployment with 5 devices (represented as red dots). The black lines represent the Voronoi polygon corresponding to each IIoTD, while the blue lines represent the sensing area border. (left) Voronoi-(i), (center) Voronoi-(ii), and (right) Voronoi-(iii).}
 \vspace{-3mm}
	\label{voronoi1}
\end{figure*}

\section{{Heuristic \& Data-driven proposed solutions}}

%\vspace{-2mm}

%The accuracy of the approximation in solving the problem in \eqref{P2} depends on several factors such as the complexity of the problem and the chosen approximation for the optimization algorithm. However, 
Relying on previous optimization techniques can be computationally expensive and time-consuming, especially for large-scale IoT systems. Herein, we aim to explore heuristic and data-driven approaches to efficiently solve the problem. 
Note that heuristic approaches may lead to near-optimal solutions relatively quickly. In contrast, data-driven solutions can learn from the available data and make predictions based on the learned patterns. 
%These approaches can be particularly useful when dealing with large-scale IoT systems. Our goal is to provide a range of solutions that are both computationally efficient and accurate in solving the problem in \eqref{P2}.
Metaheuristic methods such as GA~\cite{saraswat2013genetic} and PSO~\cite{pso1} are valid tools, along with online methods like RL~\cite{bertsekas2019reinforcement} and Lyapunov optimization~\cite{neely2022stochastic}, which also offer adaptability. 
%However, these methods come with challenges such as substantial data and potential sensor resource consumption, making them less practical in some applications~\cite{chatzieleftheriou2022online}.

In the following, we present potential approaches to solve the problem. The first two solutions are heuristics, one based on the Voronoi diagram and the other on clustering and Bayes' theorem, specifically KNN. The third solution uses metaheuristic numerical methods and requires GA and PSO tools. 
The fourth and final approach is a RL-based solution.
Notice that, unlike the metaheuristic methods, such heuristic learning can be performed online, as it does not require an oracle view of the network for re-training. Specifically, the sensors can independently implement and change strategies in case of re-training, using acknowledgments for correctly transmitted packets as their only feedback while knowing the position of the sensors (fixed) in its cluster and their threshold.

\subsection{Voronoi-based approach}

%Solving the problem in \eqref{P2} is non-trivial and using approximation methods like SCA or BCD is still not optimum and far not explaining the reason to choosing the solution. 

Herein, we optimize $\boldsymbol{\delta}$ using the Voronoi graph theory applied to the IIoT deployment. The principle of the Voronoi diagram has been maturely applied to computer graphics research, occupying an important role in computational geometry~\cite{okabe2009spatial}. A Voronoi diagram comprises a set of continuous polygons formed by vertical bisectors connecting two neighboring edges~\cite{Voronoi_ma2021escvad}. The bisector is the trajectory of all points at equal distances to neighboring IIoTD. A Voronoi diagram has three basic properties: (i) each Voronoi area is unique; (ii) the adjacent Voronoi area of each Voronoi area is the nearest adjacent area in the Euclidean plane; and (iii) each Voronoi area has at least three edges, and the edges are closed~\cite{Voronoi_ma2021escvad}. Fig.~\ref{voronoi1} depicts a simplified and easily interpretable Voronoi diagram of an IIoT deployment with 5 devices. % represented as red dots. The diagram is made up of green lines that depict the Voronoi polygon of each IIoT.

The Voronoi diagram's definition and unique properties can be harnessed to partition areas and employ the resulting polygons for each device as a metric for optimizing transmission thresholds. Specifically, we use the polygon metrics to calculate the optimal $\delta_j = e^{-\eta \Omega_j}$, where $\Omega_j$ (blue lines in Fig.~\ref{voronoi1}) represents the target coverage radio per IoTD, and it is obtained by using three approaches as follows
\begin{itemize}
    \item[i)] minimum distance from the IIoTD to its polygon;
    \item[ii)] mean distance from the IIoTD to its polygon;
    \item[iii)] maximum distance from the IIoTD to its polygon.
\end{itemize}
When employing a Voronoi-based approach, the worst-case complexity is $\mathcal{O}(2N\log N)$~\cite{liu2011output}. %However, the computation time varies based on the neighbor calculation algorithm used (such as KNN)~\cite{liu2011output}. 

\subsection{Bayesian-based KNN}

Recently, ML capabilities and utilization have tremendously increased in various fields. 
%However, this rapid evolution has led to a significant problem: the difficulty of understanding the AI and ML models. 
While cutting-edge ML models provide invaluable benefits, they often function as black boxes, which makes it difficult for humans to understand their decision-making processes. 
%The increasing popularity of complex models, particularly Deep Neural Networks (DNNs), has led to an increase in this issue. These models have a million parameters and intricate structures, they're considered to be complex black boxes, which further diminishes their understandability. 
Indeed, complex models such as deep neural networks have multiple parameters and intricate structures to the point of being considered black boxes with low understandability.
%As AI systems become more autonomous and influential in fields like healthcare, law, and defense, a growing number of people are seeking to understand and justify the decisions made by these methods.

\begin{figure}[t!]
	\centering
	\includegraphics[width=0.7\columnwidth]{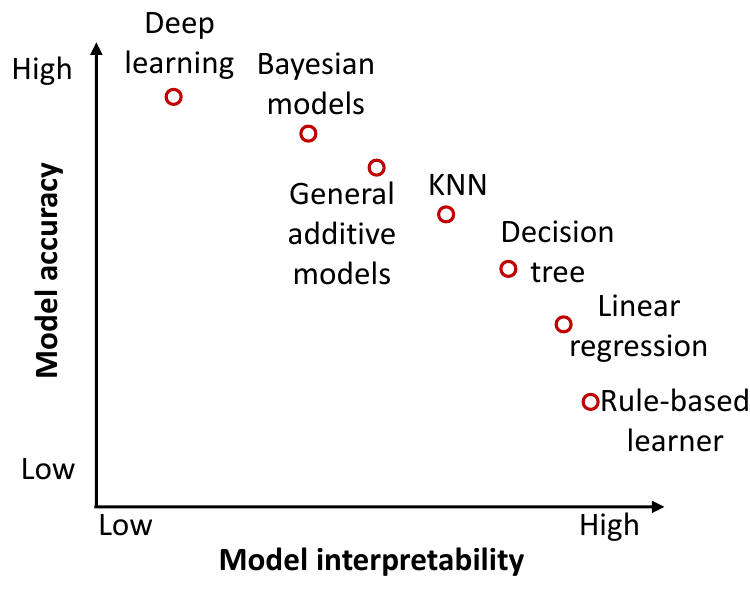}
		\vspace{-3mm}
	\caption{The trade-off between model interpretability and performance.}\vspace{-2mm}
	\label{xai}
 \vspace{-2mm}
\end{figure}

The demand for transparency has prompted the emergence of Explainable AI (XAI). %a field of study dedicated to developing ML methods that have the dual goal of producing more understandable models while still maintaining high learning efficiency, and allowing humans to trust, understand, and manage these AI systems. 
XAI is a field dedicated to developing AI methods that serve two primary objectives: i) to generate more comprehensible models while ensuring their learning efficiency remains high and ii) to enable humans to trust, understand, and effectively manage these AI systems. 
%However, when existing ML models fail to achieve transparency, post-hoc explainability methods become necessary. These methods seek to explain how a model makes predictions for a given set of inputs, thereby connecting complex models to human understanding~\cite{arrieta2020explainable}.
%To ensure the practical application of AI models and address the lack of transparency in modern ML models, XAI focuses on increasing transparency, trust, causality, transferability, informativeness, confidence, fairness, accessibility, and interactivity. This involves creating transparent ML models, 
Note, for instance, that ML models 
such as linear/logistic regression, decision trees, KNN, rule-based learners, general additive models, and Bayesian models are more easily understood and manipulated by humans. %The field of XAI has been created, its purpose is to make AI models more understandable while still maintaining their high accuracy in predictions. XAI attempts to facilitate the understanding, trust, and control of these complex AI systems.
These models are transparent by design and have a greater explaining capacity than complex models like deep neural networks, as depicted in Fig.~\ref{xai}, inspired in~\cite{arrieta2020explainable}.

Herein, we adopt KNN~\cite{KNN1,KNN2} and Bayesian Models~\cite{bayes1}, widely used XAI approaches in the IIoT context because of their trade-off accuracy/interpretability and simplicity.  
%We will discuss how using KNN and Bayes' model we can assure a good performance solution for \eqref{P2} while achieving transparency and provide information about the way AI is conducted. XAI provides a promising approach to utilizing the full potential of AI systems by making them more accessible and interpretable, this will ensure that they are both responsible and effective in various domains. 
%We introduce a heuristic where we construct the clusters around each device, grouping the devices according to their spatial deployment and neighbors' angle, \textit{i.e.}, the distance and direction of each neighbor. Note that for IIoT the network is stationary so the position of the devices is well known and can be implemented offline.
Specifically, we propose a heuristic approach wherein we form clusters based on each device's spatial placement using KNN, thus each data point is connected to its K-nearest neighbors, creating a graph structure.  
Note that the considered IIoT network is stationary, \textit{i.e.}, devices' positions are fixed and known, thus the computational cost associated with implementing KNN is $\mathcal{O}(N(N-1)/2)$ in the worst-case scenario (brute force). This complexity arises due to the need to calculate distances to the $N$ IIoTD for each query point. The search involves scanning the entire dataset to identify the KNN. However, as previously mentioned, the computation time can differ based on the utilized algorithm, occasionally reducing to $\mathcal{O}(\min \{M(N - M), (N - M)^2\})$, where $M$ is the number of clusters~\cite{liu2011output}. Note that we only form clusters once for a given IIoTD deployment, thus it can be implemented offline.

\begin{figure}[t!]
	\centering
	\includegraphics[width=0.6\columnwidth]{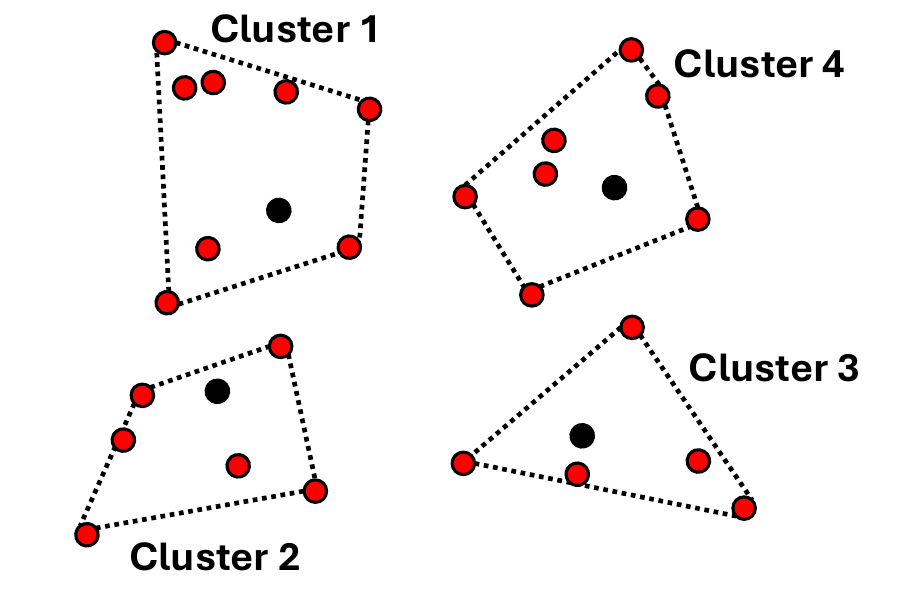}
		\vspace{-3mm}
	\caption{Illustration of an IIoT deployment clustering with 30 IIoTDs and 4 clusters. The dot lines represent the cluster edges and the black dots represent IIoTDs acting as cluster heads. }\vspace{-4mm}
	\label{knn}
\end{figure}

%Once the clusters are formed, we implement a greedy generalized Bayes' theorem starting from the edges that statistically have more probability to need a bigger sensing area (smaller threshold) to cover the entire area and avoid miss-detection. Specifically, we form clusters for each device where this device is the center of the cluster, according to the correlation (proximity) and direction to each neighboring device. 
After establishing the clusters, as depicted in Fig.~\ref{knn} (see also  Fig.~\ref{figure1}) as an example, we employ a greedy conditional probability rule, the generalized Bayes' rule. We start from the edges that statistically require a larger sensing area, \textit{i.e.}, a lower threshold, to cover the entire area effectively and prevent miss-detection. 
Each device constitutes the center of a cluster  according to the spatial correlation to its nearest neighbors, forming a graph structure. 
Initially, the devices' activation probability is set randomly. 
Then, we begin by updating the transmission threshold for each device by taking into account its spatial correlation with its nearest neighbors. We calculate $\delta_j$ based on the transmission thresholds of the neighbors and its transmission threshold, taking into account the spatial correlation as well. Then,     
\begin{equation}
    \Pr\left( p(d_{i,j}) \geq {\delta_j}\right) =  \sum_{\forall h} \Pr(A_j | A_h) \Pr\left( p(d_{i,h}) \geq {\delta_h}\right),
\end{equation}
where $\Pr(A_j | A_h)$ is the conditional probability for the device $j$ being active given that device $h$ is active.  

\begin{theorem}\label{conditional_theorem}
Let $R = \min(x_h,y_h,H-x_h,L-y_h)$ and $r = \max(R,d_h)$, 
then 
\begin{align}\label{AjAh}
    \Pr(A_j | A_h) = 1 - \frac{\cos^{-1}\left(\frac{d_{i,h}^2 + r^2_{j,h} -\mathrm{ln^2}(\delta_j)/\eta^2 }{2d_{i,j}r_{j,h}}\right)}{2\pi - 8\cos^{-1}\left(R/r\right)},  
\end{align}
where $r_{j,h}$ is the distance between the IIoTDs $j$ and $h$, and $\varphi$ represents the angle at the event epicenter determined by the IIoTDs $j$ and $h$ positions. 
\end{theorem}
    
%\begin{proof}%[\textit{Theorem} \ref{cdf_theorem}]
    %The CDF of $z$ has a closed-form and is calculated (
\textit{Proof:} See proof in Appendix~C.

Finally, the complexity when implementing the Bayesian-based KNN algorithm is bounded by $\mathcal{O}(gN + \min \{M(N - M), (N - M)^2\})$, where $g$ is the maximum number of iterations to converge. Note that the required iterations vary depending on the initial value for $\boldsymbol{\delta}$.  

%\subsection{Metaheuristic Optimization}

%Herein, we resort to three numerical methods for solving \eqref{P2}.

\subsection{Genetic algorithm (GA)}

We employ a GA for its capacity to explore extensive solution spaces and tackle nonlinear, non-convex problems efficiently. GAs are well-suited for parallelization and  scalability~\cite{du2022multiuser} and adapt to dynamic and changing environments. Unlike black-box approaches, GAs provide flexibility and adaptability while maintaining transparency~\cite{zhao2022adaptive,allahloh2023optimizing}.
These meta-heuristic algorithms inspired by biological evolution are straightforward to construct and demand relatively modest storage. 

GAs systematically explore the state space and iteratively apply  mutation, crossover, and selection operations~\cite{saraswat2013genetic}. 
Specifically, an initial population of size $\Theta$ with potential solutions is generated first.   
%\footnote
{Different $\boldsymbol{\delta}$ initialization are generated with random threshold values.} 
Note that increasing the population size increases the probability of finding a near-optimal solution, but the execution complexity also increases.   
The best-fit potential solutions are probabilistically selected from the current population based on the objective and constraint functions of \eqref{P2}. Then, the threshold parameters for each device undergo modifications, including recombination and potential random mutations, to generate a new generation~\cite{saraswat2013genetic}. Crossover combines  activation probabilities of devices from different parent solutions, while mutation introduces random changes in these probabilities to promote exploration. %The algorithm proceeds iteratively while the population evolves through successive generations, with individuals having activation probabilities adjusted based on their fitness until an optimal or near-optimal solution that satisfies the optimization objective is found or the number of iterations parameter ($g$) reaches the preset maximum number.
The algorithm works by repeatedly going through a process of evolution, where the population changes over time and individuals are adjusted based on their fitness until an optimal or near-optimal solution is found. This process continues until either the optimization objective is met or the maximum number of iterations is reached (as defined by the parameter $g$). 

Using this meta-heuristic algorithm, we approximate a near-optimal configuration with a complexity of $\mathcal{O}(g \beta)$, where $g$ is the maximum number of iterations, and $\beta$ represents the cardinality of the candidate population vector, which depends on the parameter values $N$ and $\Theta$ as follows, $\beta = N \Theta$. Parameter fitness is assessed in each generation using the objective and constraint functions in \eqref{P2}. 
%The goal is to minimize power consumption while ensuring error probability remains below a specified threshold. 

\subsection{Particle swarm optimization (PSO)}

%PSO algorithm presents a compelling choice for addressing non-convex optimization challenge in P2. 
PSO is exceptional in dynamic and nonlinear spaces with a high complexity degree~\cite{pso1} and thus it is a compelling choice for addressing P2.  
%Additionally, its flexibility to adapt to discrete or combinatorial scenarios, increases the application range in the IIoT. 
PSO operates on the principles of swarm intelligence, mirroring the collaborative behavior of particles in nature. 
This is appealing in IIoT systems, where numerous devices must work together efficiently~\cite{pso1,pso2}.

%The ability of PSO to converge to a global solution by sharing information between particles is particularly beneficial in complex IIoT environments with a high degree of intrigue. This ensures that the optimization process reduces energy consumption but maintains high fidelity. The efficiency, versatility, and capacity to deal with complex situations make PSO an ideal choice for optimizing energy efficiency in the IIoT~\cite{pso3}.
Like in GA, a population $\Theta$ of threshold sets is randomly initialized within the threshold space $(0,1]$, where each threshold set represents a potential solution. Then, we evaluate the potential solutions using \eqref{P2} and based on the optimization goal. In each iteration, the threshold sets adjust their activation probabilities based on knowledge of better individual and global solutions discovered by themselves and the swarm. 
Specifically, we determine ${\delta_j}$ for each device by using a fixed value for the other devices' thresholds and selecting the best potential solutions in the population. 
At each iteration, we assess the fitness of each particle's solution using \eqref{P2}, then update the individual ${\delta_j}$ and global best $\boldsymbol{\delta}$ based on any improvements in fitness. This allows us to refine and improve the overall solution.  
PSO continues iterations until an optimal or near-optimal solution that satisfies the optimization objective is found, or the number of iterations parameter ($g$) reaches the preset maximum number. The best global solution found represents optimized activation probabilities. 
Herein, we approximate a near-optimal configuration using PSO with a complexity of $\mathcal{O}(gN\Theta)$. Note that the population of potential solutions can be initialized based on the spatial correlation relation from heuristics like Voronoi, which may decrease the complexity to reach a near-optimal solution.  

%Furthermore, PSO offers practical advantages crucial for IoT optimization. It excels in complex, dynamic, and non-linear problem spaces, making it well-suited for the inherently unpredictable and noisy nature of IoT networks. Its adaptability to discrete and combinatorial optimization scenarios further extends its applicability in IoT, where decisions about device activation probabilities involve discrete choices. PSO's ability to converge towards a global solution by sharing information among particles is particularly valuable in intricate IoT network environments, ensuring that the optimization process effectively reduces energy consumption while maintaining stringent error probability constraints. In essence, PSO's efficiency, adaptability, and ability to handle complex scenarios make it an optimal choice for IoT energy optimization.

%\textbf{\textit{{3) Fmincon}}}

%\noindent In addition we also use MATLAB non-linear solver fmincon as benchmark for maximizing energy efficiency in the network, following the objective and constraint functons in \eqref{P2}.

\subsection{Reinforcement Learning (RL)}

Model-free RL is a programming tool to tackle decision-making challenges and learn optimal solutions in dynamic environments~\cite{RL_1}. Herein, we reframe the optimization problem by introducing spatially-based transmission thresholds and approach it as RL challenge. The IIoT system is conceptualized as the environment, while the devices function as learning agents in this setup. A coordinator retains decision-making authority to prevent computational overload at the IIoTDs, representing them virtually. The central controller stationed at the coordinator provides notifications for miss-detection and collisions. The fundamental components of RL are described as follows.

Let $\mathcal{S}$ denote the system state space. The current system state $s \in \mathcal{S}$ includes the state of each device (active or inactive) and their received sensing power from the event, depicted by the sensing function $p(d_{i,j})$. Consequently, it also includes information corresponding to collisions and miss-detections. In addition, the known position of the devices and their current thresholds are also known. The current state of each device is defined as
\begin{equation}
    \begin{array}{rl}
        s = \{ \{\delta_j\}_{\forall j\in \mathcal{J}}, \{p(d_{i,j})\}_{\forall j\in \mathcal{J}}, \{(x_j,y_j)\}_{\forall j\in \mathcal{J}} \}. 
    \end{array}
\end{equation}
%where $(x_j,y_j)$ is the spatial position of the $j^{th}$ IIoTD.
With an observed state `$s$' given, the coordinator varies $\delta$ for each device according to the reward/penalty function. The action space is then delimited by threshold limits as $\delta_j \in (0;1];$ $\forall j \in \mathcal{J}$. 
The reward must capture the effectiveness of the threshold policy when the agent takes an action in the current state. %At each learning step, the system performance follows the desired objective~\cite{RL_1}. 
%Thus, a reward function able to increase energy efficiency without incurring errors is crucial. The reward function represents the optimization goal, and our objective is to maximize system energy efficiency.  %by maintaining the activation probability in the network as close to $1/N$ as possible. 
In each learning step, the system's performance must align with the reward function~\cite{RL_1}, enhancing energy efficiency while avoiding errors. This reward function represents our optimization goal, which is to maximize the system's overall energy efficiency. 
Therefore, the reward function at a TTI $\tau$ is expressed as 
\begin{align}\label{reward}
        \!r_\tau\! =\!\!\!\!\sum_{j,h \in \mathcal{J}}\! \!\!\Pr(A_j|A_h)\!\Pr\!\left( p(d_{i,j})\! \!\geq\! \!{\delta_h}\right)\! - \!\mu_1 %\sum_{j \in N} 
        p(d_{i,j}) \rho_j\! %+ \dots \nonumber\\ 
        %& 
        - \!\mu_2 p(d_{i,j}) \sigma\!,
\end{align}
where the coefficients $\mu_1$ and $\mu_2$ are  the positive constants used to balance the utility and cost. Additionally, $\Pr(A_j|A_h)$ accounts for the conditional activation probability between IIoTD in position ($x_j,y_j$) given an active IIoTD in ($x_h,y_h$). %, which are weighted beneficially in~\eqref{reward}. 
On the other hand, the second and last terms are penalized to account for situations that provoke error. Specifically, $\rho_j$ accounts for the collision effect factor given by 
\begin{equation}
    \begin{array}{rl}
        \rho_j &= \begin{cases}
        1, &\text{if collision occurs and} \Pr\left( p(d_{i,j}) \geq {\delta_j}\right),
        \\
        \hfil -1, &\text{otherwise.}\\
    \end{cases}\\
    \end{array}
\end{equation}
Moreover, $\sigma$ is the miss-detection factor  given by 
\begin{equation}
    \begin{array}{rl}
        \sigma &= \begin{cases}
        1,   &\text{if no IIoTD is active,}\\   
        \hfil 0,   &\text{otherwise.}\\
    \end{cases}\\
    \end{array}
\end{equation}
Herein, we assume that the coordinator detects event miss-detection during training. 
Note that $\rho_j$ and $\sigma$ impose the error probability satisfaction level. If no collision or miss-detection occurs, then $\rho_j = 0$ or $\sigma = 0$, indicating no penalization to the reward function due to any error. 
The idea is finding an optimal policy $\pi^*$ at each state $\tau$ (mapping states in $\mathcal{S}$ to the probability of choosing an action %$\pi(s): S \rightarrow \varphi$)
${\pi_j^*(s) \in [-\delta_j^*,1-\delta_j^*]}$, with ${s_{\tau} + \pi^* \rightarrow s_{\tau+1}}$%, for each $j \in \mathcal{J}$
)~\cite{bertsekas2019reinforcement} that maximizes the long-term expected discounted reward. The cumulative discounted reward is given by 
\begin{equation}
    \begin{array}{rl}
        U_{\tau} = \sum_{k = 0}^{\tau} \zeta^k r_{k+1}, 
    \end{array}
\end{equation}     
where $\zeta \in (0,1]$ is the discount factor that grows exponentially with each state $\tau$ and $r_{\tau+1}$ is the reward at the next state.  

Moreover, the state-action function of the agent with a state-action pair $(s,a)$ under a policy $\pi$ is given by 
\begin{equation}
    \begin{array}{rl}
        Q^\pi_{\tau}(s_\tau,a_\tau) = \mathbb{E}_\pi [U_\tau|s =s_\tau,a=a_\tau],  
    \end{array}
\end{equation}  
where a conventional \textit{Q}-learning algorithm can be adapted to learn the optimal policy by updating the \textit{Q}-table using Bellman's equation to reach the optimal action-value function~\cite{RL_yang2020deep}. Furthermore, the \textit{Q}-value is updated as follows 
\begin{align}
        Q_{\tau+1}^\pi(s_\tau,a_\tau) &= (1-\omega_\tau)Q_{\tau}^\pi(s_\tau,a_\tau) + \dots \nonumber\\ 
        %\nonumber\\
        &+ \omega_\tau(r_\tau + \zeta_\tau \max_{a_\tau} Q^{\pi^*}(s_{\tau+1},a_{\tau+1})), 
\end{align}
where $\omega_\tau \in (0,1]$ is the learning rate. 

$Q$-Learning generally constructs a lookup $Q$-table $Q(s, a)$, and the agent selects actions based on an $\epsilon$-greedy policy for each learning step~\cite{RL_lu2023deep}. In the $\epsilon$-greedy policy, the agent chooses the action with the maximum $Q$-table value with probability $(1-\epsilon)$, whereas a random action is picked with probability $\epsilon$ to support exploration and avoid getting stuck at non-optimal policies~\cite{RL_1}. Once the optimal $Q$-function $Q^*(s, a)$ is achieved, the optimal policy is determined by
\begin{equation}
    \begin{array}{rl}
        \pi^* (s,a) = \arg \max_{a_\tau} Q^*(s, a). 
    \end{array}
\end{equation}

The complexity of the RL-based approach %is not straightforwardly represented by $\mathcal{O}()$ notation in terms of $\mathcal{J}$. The complexity 
might vary depending on the initial values and the algorithm implementation. However, the worst-case algorithm complexity is  $\mathcal{O}(N^2)$.

\section{Complexity Analysis}

Herein, we summarize and discuss the complexity of the proposed optimization algorithms. 
%When we compare the proposed approaches with the conventional deployment where each device has the same threshold (sensibility) we see the improvement in our proposal. 
We compare our proposals with a conventional implementation where each device has the same threshold (sensitivity) configured~\cite{thomsen2017traffic}, \textit{i.e.}, $\delta_j = \delta$, $\forall j \in \mathcal{J}$. 
Notice that 
the problem in \eqref{P2} is easily solved for an equal activation threshold
%for equal activation threshold, the problem in \eqref{P2} is easily solved 
since the sum and multiplication of probabilities in the objective and constraint function become an arithmetic and geometric mean with only one variable (see Appendix D). 

The complexity for equal-${\delta}$, SCA, and BCD depends on the complexity of performing IPM to solve $f({\boldsymbol{\delta}})$~\cite{colombo2007advances}, which depends on the inequality constraints and the complexity of a typically polynomial method. In this case, there is just one inequality constraint and the worst-case complexity to solve the problem using IPM is $\mathcal{O}(N^3)$ and typically converges for $g = \sqrt{N}$ iterations. Note that this value varies depending on the suitability of the initial values. Here, for equal-${\delta}$ and BCD, the problem is evaluated in a summation of $N$ terms with one variable, thus the complexity is $\mathcal{O}(N^3)$ for equal ${\boldsymbol{\delta}}$ while for BCD is 
$\mathcal{O}(N^{9/2})$ since the problem is solved $N$ times and converges in $\sqrt{N}$. Meanwhile, the complexity for SCA is $\mathcal{O}(N^{7/2})$ since there is a summation of $N$ terms with a vector ${\boldsymbol{\delta}}$ of $N$ variables. Without losing generality, we assume that the complexity of evaluating a variable is $\mathcal{O}(1)$.  

Table~\ref{complex} summarizes the methods' complexity. 
%From Table~\ref{complex} 
We can see that the equal-${\delta}$ and 
Voronoi approaches do not depend on the iterations to converge, while the rest do. 
%Thus, we can obtain a closed-form complexity, 
However, the rest depends on the specific algorithms and how fast they converge. 
Moreover, the computational complexity of implementing the proposed methods exhibits polynomial behavior. %Among these methods, 
BCD is the most complex method, while Voronoi and KNN %, and the metaheuristic approaches (GA and PSO) 
are comparatively less complex.  
Additionally, although RL is the most energy-efficient method, it is the only one that requires online implementation or a quite wide dataset. 
It is noteworthy that when implementing RL with an initial $\boldsymbol{\delta}$ based on previous algorithm results like SCA or Voronoi, rather than random $\boldsymbol{\delta}$ initialization, the complexity decreases up to $\mathcal{O}(N\log N)$. Therefore, RL could be implemented either offline with a history dataset or to retrain the system in a dynamic scenario.           

{\begin{center}
    \begin{table}[t!]
    \caption{Complexity analysis}
    \label{complex}
    \centering
    \begin{tabular}{ll}
    \hline
    {\textbf{Algorithm}} &   {\textbf{Complexity}}\\       
    \hline
    {Equal $\boldsymbol{\delta}$}    &   {$\mathcal{O}(N^3)$}\\
    {SCA}			    &	%{$\mathcal{O}(g(N^3+N^3)) \rightarrow 
    $\mathcal{O}(N^3\sqrt{N})$\\
    {BCD}               &   {$%\mathcal{O}(g(1+N^3)N) \rightarrow 
    \mathcal{O}(N^4 \sqrt{N})$}\\
    {Voronoi}           &   {$\mathcal{O}(N\log N)$}\\
    {Bayesian-based KNN}&   {$\mathcal{O}(gN + (N - M)(\min \{M, N - M\}))$}\\
    {GA}                &   {$\mathcal{O}(g N \Theta)$*}\\
    {PSO}               &   {$\mathcal{O}(g N \Theta)$}\\
    {RL}                &   {$\mathcal{O}(N^2)$}\\
    \hline
\end{tabular}
\\
\vspace{1ex}
\small{*Note that good choices for $g$ and $\Theta$ may increase with $N$.}
\vspace{-3mm}
\label{complex}
\end{table}
\end{center}

\vspace{-8mm}
\section{Results}

Consider a 50$\times$50 m$^2$ area with $N \in %[10^{-2} 10^{-1}]$ devices/m$^2$ (between 25 and 250 devices) 
[25 \text{ } 250]$ devices. 
We choose $N \geq 25$ since for smaller numbers the optimization problem often becomes unfeasible as there are not enough devices  to cover the area. Indeed, miss-detection events are very common for $N < 25$. %Additionally, ${\Pr(A_j | A_h) = 0.56\times e^{-0.73r_{j,h}}}$ from fitting the values in~\cite{ul2022learning}, being $r_{j,h}$ the distance between both devices. 
Additionally, we consider $\alpha = 0.1$,  %non-bursty traffic (${q = 0}$), and 
${\eta=1}$~\cite{FWuS}, and set $E=0.1$ and $g=\sqrt{N}$. 
%\textcolor{blue}{Table~\ref{setup} 
%summarizes the parameters used in the simulation process.} 
We perform 250 Monte Carlo runs corresponding to different deployments. 
For simplicity, the power consumption is given without units (percentage of time in active state). However, the value can be obtained by multiplying this value and the power consumption of the specific device in an active state.
%{\begin{center}
    %\begin{table}[t!]
    %\caption{Simulation Parameters}
    %\label{setup}
    %\centering
    %\begin{tabular}{ll}
    %\hline
    %{\textbf{Parameter}} &   {\textbf{Value}}\\       
    %\hline
    %{$\xi$}    &    {50$\times$50 m$^2$}\\
    %{N}		   &	$[25 \text{ }250]$\\
    %{$\alpha$} &    {0.1}\\
    %{$\eta$}   &    {1}\\
    %{$E$}      &    0.1\\
    %{$\delta$} &    $[10^{-4}, 1]$\\
    %\hline
%\end{tabular}
%\\
%\vspace{1ex}
%\small{*Note that good choices for $g$ and $\beta$ may increase with $N$.}
%\label{setup}
%\end{table}
%\end{center}

\subsection{Benchmark and Voronoi approaches}\label{benchmark}

We adopt the equal-${\delta}$ approach as a benchmark.
Initially, we assess the power consumption linked to employing the equal-${\delta}$ approach and compare it with the four Voronoi-based solutions. It's noteworthy that the optimization problem might not always be solvable with the equal-${\delta}$ approach and the latter three Voronoi-based alternatives, due to non-compliance with the error constraints. 
%First of all, we evaluate the power consumption when using the equal $\delta$ approach and compare it with the four Voronoi-based solutions since the equal $\delta$ approach and the last three options for the Voronoi-based approach a solution for the optimization problem is not always feasible since the error constraints are not met. 

In Fig.~\ref{voronoi2}, we show the mean energy consumption per device in the network when using the two previous solutions. 
Note that the Voronoi-(i) algorithm has the best performance, while the scenario with equal-${\delta}$ and Voronoi-(iii) has the worst performance related to energy consumption. Herein, note that equal-${\delta}$ slightly outperforms Voronoi-(iii) for more than 33 IIoTDs.    
Indeed, the sensing area in the latter approach is determined by the biggest distance to the Voronoi polygon, hence, the overlapping probability in sensing areas for neighbor IIoTD is high and the energy efficiency low. 
However, Voronoi-(i) is the only algorithm capable of finding a solution to the optimization problem in each scenario. The other approaches can find solutions in just some cases, as depicted in Fig.~\ref{feasible}. In this case, both approaches with the worst performance can find a solution between 49\% (equal-${\delta}$) to 55\%  (Voronoi-(iii)) of the time, while the feasibility is about 72\%-74\% for Voronoi-(ii).   

\begin{figure}[t]
	\centering
	\includegraphics[width=\columnwidth]{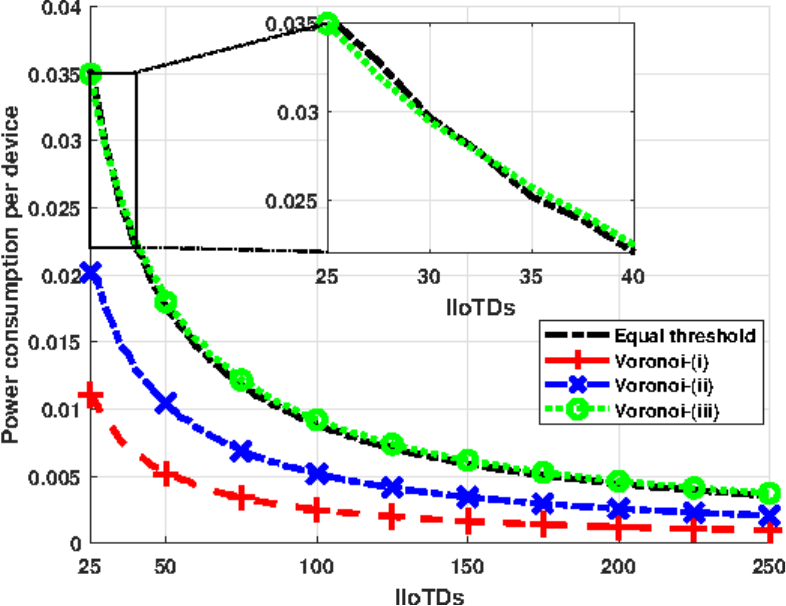}\vspace{-3mm}    \caption{Power consumption as a function of the number of IIoTDs for the benchmark and the Voronoi approaches.}\vspace{-2mm}
	\label{voronoi2}
\end{figure}
\begin{figure}[t]
	\centering
	\includegraphics[width=\columnwidth]{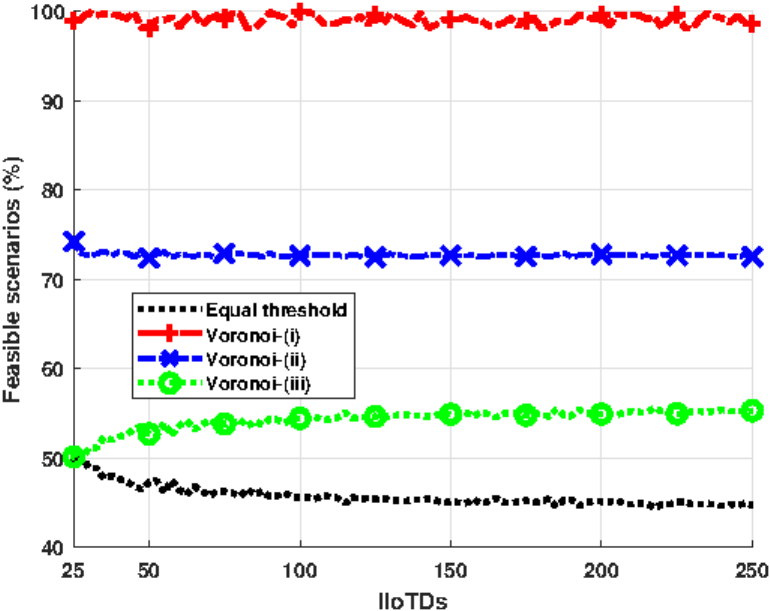}
 \vspace{-7mm}
    \caption{Feasibility rate to find a configuration with $P_e \leq 0.1$ as a function of the number of IIoTDs for the benchmark and Voronoi's approaches.}
	\vspace{-2mm}\label{feasible}
\end{figure}

Based on the above discussions, only the Voronoi-(i) approach is adopted for comparison purposes in the following. %This is because it is the only approach that meets the 100\% feasibility constraint, which is a requirement for all solution proposals in the following section. 
It is worth noting that the equal-${\delta}$ solution, although not always feasible according to the constraint, will still be included in the comparisons for benchmarking purposes.

\subsection{Performance comparison}

Herein, we present performance results for the proposed energy-efficient solutions. Notably, these solutions are feasible across all 250 modeled scenarios, unlike the benchmark and Voronoi's approaches analyzed in Section~\ref{benchmark}. Specifically, Fig.~\ref{error1} shows the power consumption per device as a function of the device density and using the approximation methods. 
Herein, BCD and RL have the worst and best performance, respectively. Interestingly, as we increase the device density, SCA and KNN show a performance similar to RL. Meanwhile, GA and PSO show similar performances and outperform the equal-${\delta}$ approach. On the other hand, BCD outperforms the equal threshold scheme for a density below 150 IIoTDs. However, it consumes more power when the density increases above 150. 
Notice that for the equal-${\delta}$ scenario, the error probability constraints can not be met for around half of the scenarios modeled. Here, the RL approach reduces the power consumption by up to 96\% compared with the benchmark for low-density scenarios,  while for high-density scenarios, the consumption is reduced by 60\%. As expected, the power consumption decreases as IIoTDs density increases because coverage increases, allowing for a smaller sensing area to be set for each IIoTD, thereby lowering their activation probability for most approaches.    

%We calculate, in Fig.~\ref{error1}, the relative energy consumption using the approximation methods regarding the case when we assume that each device has equal activation threshold. We can see how both approaches outperform the benchmark with energy savings up to 40 and 85\% for BCD y SCA respectively.  
%Notice that SCA performs better than BCD  when the device density goes from 0.04 to 0.1 while for values between 0.02-0.04 is BCD the one with the best performance. However, despite similar performance below a density equal to 0.04, while we increase the density the gap between SCA and BCD is notable. In this case, SCA is capable of reach a better solution. 

\begin{figure}[t]
	\centering
	\includegraphics[width=\columnwidth]{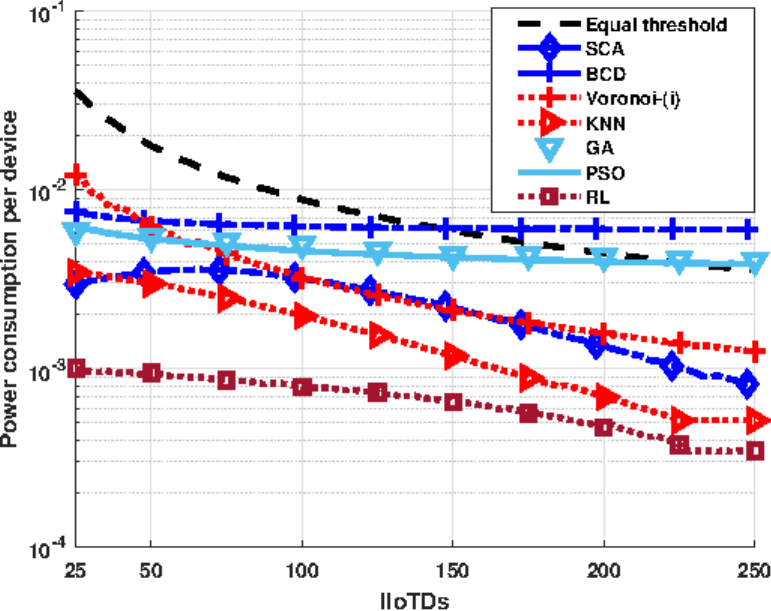}
 \vspace{-7mm}
    \caption{Power consumption as a function of the number of IIoTDs per device for the proposed solutions.}
    \vspace{-2mm}
	\label{error1}
\end{figure}

\begin{figure*}[t!]
    \centering
    \includegraphics[width=0.65\columnwidth]{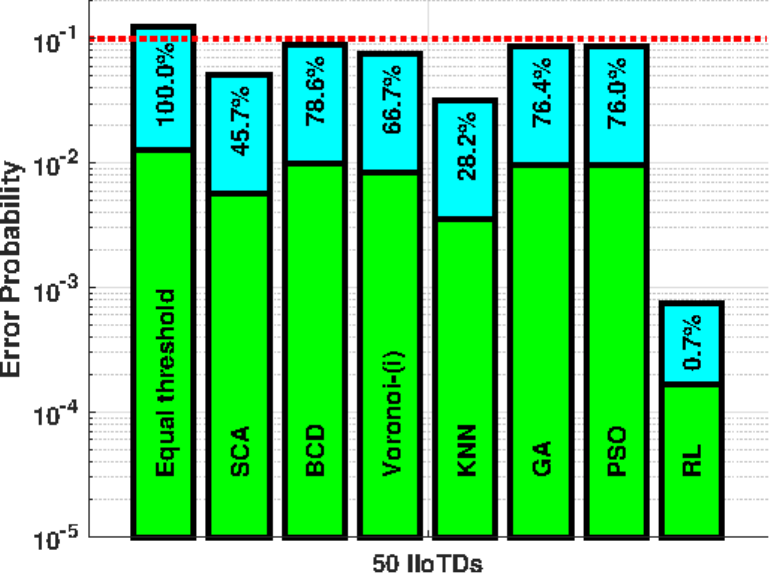}
    \includegraphics[width=0.65\columnwidth]{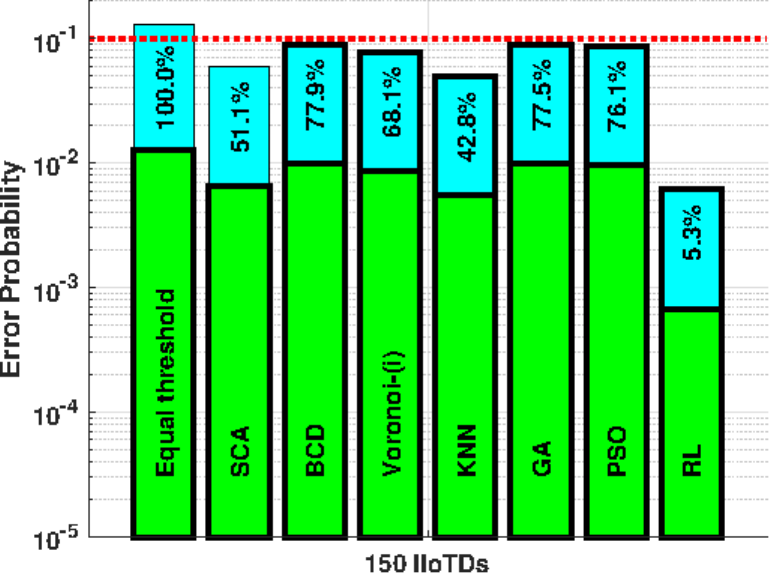}
    \includegraphics[width=0.65\columnwidth]{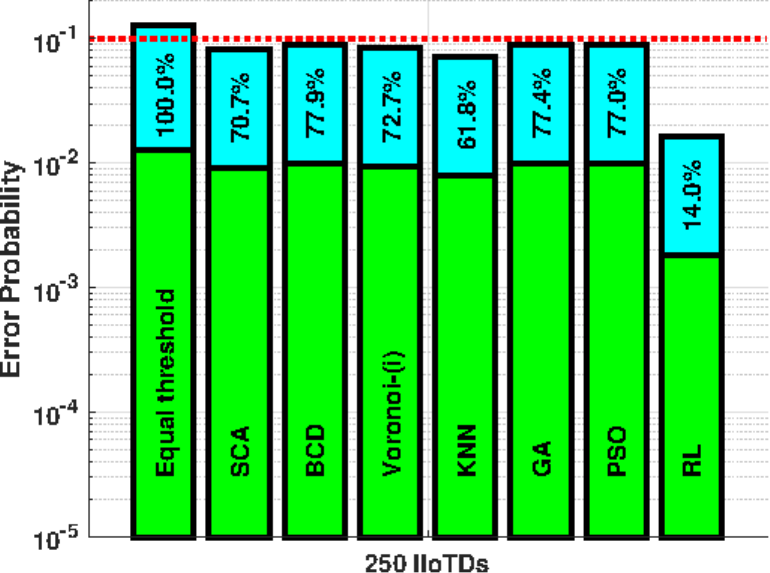}
    \vspace{-2mm}
\caption{Error probability ($P_{\text{e}}$) for the proposed solutions as the sum of the collision (cyan) and miss-detection (green) probabilities for (left) 50, (center) 150, and (right) 250 IIoTDs. The percentages indicate the error probability of the methods compared to the equal-$\delta$ approach, while the red dotted line represents the error constraint $E$.}
    \label{error3}
\end{figure*} 

In addition, Fig.~\ref{error3} displays  $P_{\text{e}}$ for the proposed methods, which combines the collision (cyan) and miss-detection (green) probabilities. It's worth noting that $P_{\text{e}}$ is consistently below 0.1 for each of the proposed solutions, meeting the constraint $E$ in P2, while the equal-$\delta$ approach fails to stay below this limit. The value of $P_{\text{miss}}$ remains almost constant, staying below 0.01. It's important to highlight that the RL approach demonstrates the lowest $P_{\text{e}}$, whereas BCD, GA, and PSO exhibit the highest values. Notably, the value of $P_{\text{e}}$ increases with higher IIoTD density, as more devices are likely to collide upon activation from the same event. Furthermore, Fig.~\ref{error3} also  illustrates the relative error probability of the proposed methods compared to the equal-$\delta$ approach. Notably, in low-density scenarios, SCA, KNN, and RL reduce the error probability by up to 45.7\%, 28.2\%, and 0.7\% respectively. However, in high-density scenarios, only RL is capable of reducing the error probability by up to 14\%, while the other approaches decrease the error probability by around 61\%-77\% compared to the equal-$\delta$ approach.    

Noteworthy, employing BCD requires a computational cost $N$ times higher than that of SCA. This is even though each iteration in SCA is slightly more complex to execute. Furthermore, RL requires access to additional historical data characterizing the IIoTD traffic and activation behavior to facilitate learning, which makes the training a little more complex but gives additional online support for a dynamic scenario. 
%Also, as the scenario exhibits characteristics resembling clusters, the performance of the proposed methods can be further enhanced. 

%Notice that varying $\eta$ has no effect in power consumption  for our proposed methods since the sensing area (threshold) is set according to $\eta$, therefore the threshold value varies accordingly but not the sensing area. On the other hand, increasing the value of $q$ might increase the power consumption since more time is needed in active state to transmit information about a single event, this increment is associated with the expected mean value of q distribution function, $q/(1-q)$. However, the impact in collisions is negligible for this scenario but for a high traffic rate scenario might incur in collisions due to IIoTDs trying to transmit information of different overlap events, which is not consider in this paper. 
Note that the value of $\eta$ does not affect the power consumption of our proposed methods. This is because the transmission threshold is adjusted based on the value of $\eta$, but the sensor area itself remains constant. %On the other hand, increasing the value of $q$ may result in higher power consumption\footnote{The increment is associated with the expected mean value of the distribution function of $q$, calculated as $\frac{q}{1-q}.$} since it requires more time in the active state to transmit information about a single event. The impact on collisions, however, is negligible in this scenario. However, for high traffic rate scenarios, there may be collisions due to IIoTDs attempting to transmit information about different overlapping events, which is beyond the scope of this paper.

\section{Conclusion}

This paper comprehensively explored energy efficiency and transmit resource allocation strategies in IIoT setups while considering device-specific attributes and activation correlations. 
We introduced a transmission threshold for event-sensing scenarios and assessed its effects on energy efficiency, coverage area, and successful event detection. 
We formulated the optimization problem for setting the transmission threshold using a convex approximation. Likewise, we presented multiple solutions relying on convex methods such as SCA and BCD; and heuristic methods like Voronoi diagrams, explainable ML, and algorithms based on natural selection and social behavior. 
Additionally, we reformulated the problem as a form of RL, employing $Q$-learning. We showed that the complexity of the proposed methods is polynomial, with BCD as the most complex, while Voronoi and KNN are the least complex. 
Overall, our proposals provided a 94\% power consumption reduction concerning the equal-$\delta$ benchmark in low-density scenarios, while the consumption is reduced by 60\% for high-density scenarios. %Additionally, the complexity for performing the proposed methods has polynomial complexity with BCD being the more complex while Voronoi, KNN, and the metaheuristics (GA and PSO) have the less complex algorithm. 

In future work, we aim to explore adaptive methods for optimizing transmission thresholds dynamically according to network conditions, non-stationary devices, or device-specific attributes. This could involve developing algorithms that adjust thresholds in real-time to maximize energy efficiency while maintaining a target performance.

\section*{Appendix A. CDF of $d_{i,j}$}\label{append1}

%\begin{proof}
The variable $(X-x_j)$ has a uniform probability distribution function (PDF) $ \textstyle \frac{1}{L} \in [-x_j, L-x_j]$ and the PDF of 
${\hat{X} \triangleq (X- x_j)^2}$   
is given by~\cite{miller2012probability}
\begin{align}
        f_{\hat{X}}(x) 
        &= \begin{cases}
                        \frac{1}{u\sqrt{\hat{X}}},   &0 \leq \hat{X} \leq \text{min}(x_j;L-x_j)^2, \vspace{1mm}\\
                        \frac{1}{2u\sqrt{\hat{X}}},   &\text{min}(x_j;L-x_j)^2 < \hat{X} \leq 
                        u,\\   
                        \hfil 0,   &\text{otherwise,}\\
                        \end{cases}
\end{align}
where $u = \text{max}(x_j;L-x_j)^2$. Likewise, let us denote ${\hat{Y} \triangleq (Y-y_j)^2}$, and then 
\begin{align}
        f_{\hat{Y}}(y) 
        &= \begin{cases}
                        \frac{1}{v\sqrt{\hat{Y}}},    &0 \leq \hat{Y} \leq \text{min}(y_j;H-y_j)^2,\vspace{1mm}\\
                        \frac{1}{2v\sqrt{\hat{Y}}},    &\text{min}(y_j;H-y_j)^2 < \hat{Y} \leq 
                        v,\\    
                        0,   &\text{otherwise,}\\ 
                        \end{cases}
\end{align}
where $v = \text{max}(y_j;H-y_j)^2$. Next, given ${Z = d_{i,j}^2 = \hat{X} + \hat{Y}}$ and knowing that $\hat{X}$ and $\hat{Y}$ are independent variables, the CDF of the variable $Z$, denoted as $F_Z(z)$, is calculated as ${\Pr(\hat{X} + \hat{Y} \leq z)}$ 
%\begin{subequations}
\begin{align}
    F_Z(z) = \int_0^u \int_0^{z-x} f_{\hat{X}}(x) f_{\hat{Y}}(y) \partial y \partial x  
    = \frac{2}{\xi}\int_0^u \frac{\sqrt{z-x}}{\sqrt{x}} \partial x.
    %F_z(z) &= \int_0^z \int_0^{w-v} f_{x,y} \partial x \partial y,\\
    %&= \frac{1}{\xi}\int_0^v \int_0^{w-v} \frac{1}{\sqrt{z_x z_y}} \partial z_x \partial z_y,
    %\\
    %&= \frac{2}{\xi}\int_0^v  \frac{\sqrt{w-v}}{\sqrt{z_y}} \partial z_y,
    %F_Z(z) = \int_0^z f_{Z}(z) \partial z = \frac{2}{\xi}\int_0^z \frac{\sqrt{w-z}}{\sqrt{z}} \partial z.
    %\\
    %&= \frac{2}{\xi}\int_0^z  \frac{\sqrt{w-v}}{\sqrt{z_y}} \partial z,
    \vspace{-3mm}
\end{align} 
%\end{subequations}
Then, using $q = \sqrt{x}$, we obtain
\begin{alignat}{4}
    F_Z(z) &= \frac{2}{\xi}\int_0^{\sqrt{x}} 2{\sqrt{z-q^2}} \partial q \nonumber\\
    &= \frac{4}{\xi}\int_0^{\sqrt{x}} {\sqrt{z\left(1-\frac{q^2}{z}\right)}} \partial q \nonumber\\
    %&= \frac{4}{\xi}\int_0^{\sqrt{z}} {\sqrt{w}\sqrt{1-\frac{u^2}{w}}} \partial u,\\
    &= \frac{4\sqrt{z}}{\xi}\int_0^{\sqrt{x}} {\sqrt{1-\frac{q^2}{z}}} \partial q.
    \vspace{-3mm}
\end{alignat}
%\vspace{-2mm}
Next, applying trigonometric substitution we obtain 
\begin{alignat}{4}\label{trigo}
    F_Z(z) &= \frac{4\sqrt{z}}{\xi}\int_0^{\arcsin(\sqrt{1/z}\sqrt{x})} \sqrt{z} \cos^2(v) \partial v \nonumber\\
    &= \frac{4{z}}{\xi}\int_0^{\arcsin(\sqrt{1/z}\sqrt{x})} \cos^2(v) \partial v\nonumber\\ 
    &= \frac{4{z}}{\xi}\int_0^{\arcsin(\sqrt{1/z}\sqrt{x})} \frac{1 + \cos(2v)}{2} \partial v,
\end{alignat}
where the last step comes from using the trigonometric identities $\cos{2\phi} + 2\sin^2{\phi} = 1$ and $\cos^2{\phi} + \sin^2{\phi} = 1$. Then, after solving the integral in \eqref{trigo}, we derive~\eqref{cdf}. \hfill$ \blacksquare $    
%\end{proof}%as follows 
%\begin{alignat}{4}
    %F_Z(z) &= \frac{4{w}}{\xi}\int_0^{\arcsin(\sqrt{1/w}\sqrt{z})} \frac{1 + \cos(2v)}{2} \partial v,\\
    %&= \frac{2{w}}{\xi}\int_0^{\arcsin(\sqrt{1/w}\sqrt{z})} {1 + \cos(2v)} \partial v,\\
    %&= \frac{2{w}}{\xi} \left( \int_0^{\arcsin(\sqrt{1/w}\sqrt{z})} {1} \partial v + \int_0^{\arcsin(\sqrt{1/w}\sqrt{z})} {\cos(2v)} \partial v \right),\\
    %&= \frac{2{w}}{\xi} \left( \arcsin{\left( \sqrt{\frac{z}{w}}\right)} + \frac{1}{2}\sin \left( 2\arcsin{\left( \sqrt{\frac{z}{w}}\right)}\right) \right).
%\end{alignat}
%Then, applying trigonometric substitution and using the trigonometric identity $\cos{2\phi} + 2\sin^2{\phi} = 1$, we obtain~\eqref{cdf}. 
%\begin{align}\label{cdf_append}
    %F_z(z) = \frac{2z}{\xi}\left( \text{arcsin}\left(\sqrt{\frac{z}{w}}\right) + \frac{1}{2}\sin{\left(2 \text{arcsin}\left(\sqrt{\frac{z}{w}}\right)\right)} \right).
%\end{align}

\section*{Appendix B. Proof of Non-Convexity of \eqref{E_Pe_convex}}\label{append2}

The first-order partial derivative of~\eqref{E_Pe_convex} is  
%\begin{subequations}
\begin{alignat}{4}\label{1partial}
        \!\nabla_{\delta_j} \mathbb{E}(P_{\text{e}}(\boldsymbol{\delta}))\! &=\!  \frac{4\alpha \mathrm{ln}(\delta_j)}{w_j \delta_j} \!\left( \sum_{h=1}^{N-1}\! e^{-2 \sum_{i\neq h}^N\! \frac{\ln^2 \delta_i}{w_i}} \!\!-\! e^{-2 \sum_{i=1}^N\! \frac{\ln^2 \delta_i}{w_i}} \!\!\right).
        %-\frac{4 {\alpha}{N}\mathrm{ln}(\delta_j) e^{-2 \sum_{i=1}^N \frac{\ln^2 \delta_i}{w_i}} }{w_j \delta_j},\\
        %&=- \alpha \sum_{h=1}^{N} e^{-2 \sum_{j\neq h}^N \frac{\ln^2 \delta_j}{w_j}} +  {\alpha}{N} e^{-2 \sum_{j=1}^N \frac{\ln^2 \delta_j}{w_j}}
    \end{alignat}
%\end{subequations}
%and similar case with $\nabla_{\delta_2} \mathbb{E}(P_{\text{miss}})$, 
Note that this expression depends on the $N$ variables, which means it is not constant. Therefore, we cannot conclude the convexity of the function based on the first-order conditions alone, and thus proceed to test the second-order condition. In this case, the Hessian matrix is given by
\begin{equation}
  \mathcal{H} =
  \left[ {\begin{array}{cc}
    h_{jj} & h_{jk} \\
    h_{kj} & h_{kk} \\
  \end{array} } \right],
\end{equation}
where $k\ne j$, ${h_{jj} = \nabla^2_{\delta_j^2} \mathbb{E}(P_{\text{e}})}$, ${h_{jk} = \nabla^2_{\delta_j\delta_k} \mathbb{E}(P_{\text{e}})}$, ${h_{kj} = \nabla^2_{\delta_k\delta_j} \mathbb{E}(P_{\text{e}})}$, and $ {h_{kk} = \nabla^2_{\delta_k^2} \mathbb{E}(P_{\text{e}})}$. Then, we have that 
\begin{subequations}\label{hjj}
\begin{alignat}{2}
        %\nabla^2_{\delta_1^2} \mathbb{E}(P_{\text{miss}}) 
        h_{jj}\! =& \frac{4\alpha}{w_j}\!\left(\!1\!-\! \frac{\mathrm{ln}(\delta_j)}{\delta_j^2}\!\right)\!\! \!\left( \sum_{h=1}^{N-1}\! e^{-2 \sum_{i\neq h}^N\! \frac{\ln^2 \delta_i}{w_i}} \!\!-\! e^{-2 \sum_{i=1}^N\! \frac{\ln^2 \delta_i}{w_i}} \!\!\right) \!\!+\! \!\dots\! \nonumber\\
        &+ \frac{16\alpha^2\mathrm{ln^2}(\delta_j)}{w_j^2\delta_j^2} e^{-2 \sum_{i=1}^N \frac{\ln^2 \delta_i}{w_i}},\\
        %-4e^{\frac{-2 \mathrm{ln^2}(\delta_2)}{w_2}} \left(\frac{-4\mathrm{ln^2}(\delta_1) + w_1 -w_1\mathrm{ln}(\delta_1)}{e^{\frac{2 \mathrm{ln^2}(\delta_1)}{w_1}} w_1^2 \delta_1^2 } \right), \\
        %\nabla^2_{\delta_2^2} \mathbb{E}(P_{\text{miss}}) 
        %h_{22} &= -4e^{\frac{-2 \mathrm{ln^2}(\delta_1)}{w_1}} \left(\frac{-4\mathrm{ln^2}(\delta_2) + w_2 -w_2\mathrm{ln}(\delta_2)}{e^{\frac{2 \mathrm{ln^2}(\delta_2)}{w_2}} w_2^2 \delta_2^2 } \right), \\
        %\nabla^2_{\delta_1\delta_2} \mathbb{E}(P_{\text{miss}}) 
        h_{jk} \!=& 16\alpha\! \left(\!\frac{\mathrm{ln}(\delta_j) \mathrm{ln}(\delta_k)}{\delta_j w_j w_k \delta_k}\! \right)\!\! \!\left( \sum_{h=1}^{N-2}\! e^{-2 \sum_{i\neq h}^N\! \frac{\ln^2 \delta_i}{w_i}} \!\!-\! e^{-2 \sum_{i=1}^N\! \frac{\ln^2 \delta_i}{w_i}}\!\! \right).
    \end{alignat}
\end{subequations}
%$\nabla^2_{\delta_2\delta_1} \mathbb{E}(P_{\text{miss}})$ 
Similarly, $h_{kk}$ is calculated by substituting $\delta_j$ and $w_j$ in (\ref{hjj}a) for $\delta_k$ and $w_k$, while $h_{kj}$ can be calculated using (\ref{hjj}b) by exchanging $\delta_k, w_k$ and $\delta_j, w_j$.
%while $h_{kk}=h_{jj}$ and $h_{kj}=h_{jk}$. 
To determine convexity, let us evaluate the Hessian at some specific values, \textit{e.g.}, $\delta_1 = 0.3$, $\delta_2 = 0.5$, and $N=2$. Then, the eigenvalues of the matrix are negative, so the Hessian is not positive semi-definite, therefore the function is not convex in the range $[0\text{ }1]$. %Similarly, the same applies to $\mathbb{E}(P_{\text{col}})$ as well. 
\hfill$ \blacksquare $

\section*{Appendix C. Closed-form for $\Pr(A_j | A_h)$}\label{append4}

%which is calculated using 
Using the cosine rule, we have  
\begin{equation}
   d_{i,j}^2 = d_{i,h}^2 + r_{j,h}^2 - 2d_{i,h}r_{j,h}\cos{\varphi}.
\end{equation}
%where $r_{j,h}$ is the distance between the IIoTDs $j$ and $h$, and $\varphi$ represents the angle at the event epicenter determined by the IIoTDs $j$ and $h$ positions.  
%between the vectors $d_{i,j}$ and $r_{j,h}$. 
Then, % is calculated as %$\Pr\left( d_{i,j}^2 \leq \mathrm{ln^2}(\delta_j)/\eta^2 \text{ }|\text{ } d_{i,j}^2 \right)$  
\begin{align}
   \Pr(A_j | A_h) =&\Pr\left( d_{i,j}^2 \leq \mathrm{ln^2}(\delta_j)/\eta^2 \text{ }|\text{ } d_{i,h}^2 \right) \nonumber\\ 
   =& \Pr\left( d_{i,h}^2 + r_{j,h}^2 - 2d_{i,h}r_{j,h}\cos{\varphi} \leq \mathrm{ln^2}(\delta_j)/\eta^2 \right),
\end{align}
where $d_{i,h}$ and $r_{j,h}$ are known. %, and $\varphi$ has a uniform distribution in $2\pi$.  
%We have that $\Pr(A_j | A_h)$ is calculated as
%\begin{equation}
   %\Pr\left( d_{i,j}^2 + r_{j,h}^2 - 2d_{i,j}r_{j,h}\cos{\varphi} \leq \mathrm{ln^2}(\delta_j)/\eta^2 \right),
%\end{equation}
%where $d_{i,j}$ and $r_{j,h}$ are known. 
Therefore,  
\begin{equation}
   \Pr\left( {\varphi} \geq \cos^{-1}\left(\frac{d_{i,h}^2 + r_{j,h}^2 -\mathrm{ln^2}(\delta_j)/\eta^2 }{2d_{i,h}r_{j,h}}\right) \right). 
\end{equation}
Let ${r = \max(R,d_h^{(i)})}$ and ${R = \min(x_h,y_h,H-x_h,L-y_h)}$. Herein, for $d_h^{(i)} \leq R$, $\varphi$ has a PDF given by $1/(2\pi)$. However, for $d_h^{(i)} > R$, the PDF varies due to the corner effect in the rectangular area. Then, to calculate the PDF taking into account the corner effect, we divided the area in 8 equal octants, $k\pi/4 \leq \varphi < (k+1)\pi/4$, for $k \in \{0,1,2,3,4,5,6,7\}$. Then, let us make a circumference with center at the expected position $(H/2,L/2)$. Focusing on the firt octant, $0 \leq \varphi < \pi/4$, the part of the circumference with $0 \leq \varphi \leq \cos^{-1}\left(R/r\right)$ fall outside the coverage area $\xi = L \times H$. Then, this values of $\varphi$ have zero occurrence probability. The same applies for the other 7 octants where $R$ coincides with $\varphi = k\pi$, for $k \in \{0,1/2,1,3/2\}$. Therefore, the PDF of $\varphi$ within $\xi$ given by
\begin{align}
        f(\varphi) 
        &\!=\! \!\begin{cases}
                        0,   &k\pi\!-\!{R}/{r}\!<\varphi\!< k\pi\!+\!{R}/{r},\\
                        \frac{1}{2\pi - 8\cos^{-1}\left(R/r\right)},   &\text{otherwise,}\\ 
                        \end{cases}%\\
                        %\\
\end{align}
for $k \in \{0,1/2,1,3/2\}$. Then, the CDF is calculated as 
\begin{equation} 
   F_{\varphi}(a) = \int_0^a f(\varphi) \partial \varphi = \frac{a}{2\pi - 8\cos^{-1}\left(R/r\right)}.
\end{equation}
%$F_{\varphi}(a) = \int_0^a f(\varphi) \partial \varphi$ and 
Then, $\Pr\left( {\varphi} \geq a \right)$  is equal to ${1 - \Pr\left( {\varphi} < a \right)}$. Herein, $\Pr\left( {\varphi} < a \right)$ can be  calculated as $F_{\varphi}(a)$. 
Therefore, the conditional probability is given in closed-form as~\eqref{AjAh}. \hfill$ \blacksquare $  
%\begin{equation}
   %\Pr\left( {\varphi} \geq \cos^{-1}\left(\frac{\mathrm{ln^2}(\delta_j)/\eta^2 - d_{i,j}^2 - r_{j,h}^2}{- 2d_{i,j}r_{j,h}}\right) \right). 
   %\Pr(A_j | A_h) = 1 - \frac{\cos^{-1}\left(\frac{\mathrm{ln^2}(\delta_j)/\eta^2 - d_{i,j}^2 - r_{j,h}^2}{- 2d_{i,j}r_{j,h}}\right)}{2\pi - 8\cos^{-1}\left(R/r\right)}.
%\end{equation}

\section*{Appendix D. Benchmark solution for P2 \eqref{P2}}\label{append3}

We rewrite \eqref{P2} assuming the equal-$\delta$ approach as follow
\begin{subequations}
\begin{alignat}{4}
        \text{P}2:\text{ }  &\min_{{\delta}} \text{ } &&{F_Z \left(\mathrm{ln^2}(\delta)\right)} \\
        &\text{ } \text{s.t.} &&\mathbb{E}(P_{\text{e}}) \leq \text{  } 0.1,
    \end{alignat}
    \label{P2_app}
\end{subequations} 
\hspace{-1.5mm}where $\mathbb{E}(P_{\text{e}})$ is given by  
\begin{align}\label{E_Pe}
        \mathbb{E}(P_{\text{e}}) = 1 - {N} (1 - e^{-\frac{2\mathrm{ln^2}(\delta)}{w}} ) e^{-\frac{2\mathrm{ln^2}(\delta)({N-1})}{w}}. 
\end{align}
%Herein, we can remove $N$ from the objective function in \eqref{P2_app} since $N$ is given and cannot be minimized. %Therefore, 
%\begin{subequations}
%\begin{alignat}{4}
        %\text{P}2:\text{ }  &\min_{{\delta_j}} \text{ } &&{F_Z \left(\mathrm{ln^2}(\delta_j)\right)} \\
        %&\text{ } \text{s.t.} &&\left(1 - e^{-\frac{2\mathrm{ln^2}(\delta)}{w}} \right) \left(e^{-\frac{2\mathrm{ln^2}(\delta)}{w}}\right)^{N-1} \geq \text{  } \frac{0.9}{N}.
    %\end{alignat}
    %\label{P2_app2}
%\end{subequations} 
%\hspace{-1.3mm}
Note that minimizing $F_Z \left(\mathrm{ln^2}(\delta)\right)$  %${1 - e^{-\frac{2\mathrm{ln^2}(\delta_j)}{w}}}$
%\begin{subequations}
%\begin{alignat}{4}
        %\text{P}2:\text{ }  &\min_{{\delta}} \text{ } &&{1 - e^{-\frac{2\mathrm{ln^2}(\delta_j)}{w}}} \\
        %&\text{ } \text{s.t.} &&\left(1 - e^{-\frac{2\mathrm{ln^2}(\delta)}{w}} \right) \left(e^{-\frac{2\mathrm{ln^2}(\delta)}{w}}\right)^{\mathcal{J}-1} \geq \text{  } \frac{0.9}{\mathcal{J}}, 
    %\end{alignat}
    %\label{P2_app3}
%\end{subequations}
%\hspace{-1.3mm}
%is equal to  maximize the second term $\exp\left({-{2\mathrm{ln^2}(\delta)}/{w}}\right)$, 
is attained by maximizing $\delta$. This is because $F_Z$ is monotonically increasing in $z$ and $z = \mathrm{ln^2}(\delta)$ is inversely proportional with respect to $\delta \in (0,1]$. 
Therefore, the solution for P2 is given by
\begin{equation}
%\begin{alignat}{4}
        %\text{P}2:\text{ }  &\max \text{ } &&{\delta_j}\\
        %&\text{ } \text{s.t.} &&\left(1 - e^{-\frac{2\mathrm{ln^2}(\delta)}{w}} \right) \left(e^{-\frac{2\mathrm{ln^2}(\delta)}{w}}\right)^{N-1} \geq \text{  } \frac{0.9}{N}, 
        \delta^\star = \sup \left\{ \delta : (1 - e^{-\frac{2\mathrm{ln^2}(\delta)}{w}} ) e^{-\frac{2\mathrm{ln^2}(\delta)({N-1})}{w}} \geq \text{  } \frac{0.9}{N} \right\},
   % \end{alignat}
    \label{P2_app4}
\end{equation}%\hspace{-1mm}
where $\delta$ is calculated as the maximum value that satisfy the constraint in P2. 

In scenarios where $N \rightarrow \infty$, as may occur in practical applications with very large $N$, an asymptotic solution emerges. In this case, the first term of \eqref{P2_app4} approaches 1, while the second term converges to 0 and becomes dominant. As a result, we can simplify \eqref{P2_app4} by solely focusing on the second term as follows
%\begin{equation}
\begin{align}
        %\text{P}2:\text{ }  &\max \text{ } &&{\delta_j}\\
        %&\text{ } \text{s.t.} &&\left(1 - e^{-\frac{2\mathrm{ln^2}(\delta)}{w}} \right) \left(e^{-\frac{2\mathrm{ln^2}(\delta)}{w}}\right)^{N-1} \geq \text{  } \frac{0.9}{N}, 
        \delta^\star &= \sup \left\{ \delta : e^{-\frac{2\mathrm{ln^2}(\delta)({N-1})}{w}} \geq \text{  } \frac{0.9}{N} \right\} \nonumber\\
        &= \displaystyle{e^{\displaystyle{\sqrt{\frac{w (\mathrm{ln}(N)-\mathrm{ln}({0.9}))}{2(N-1)}}}}}%-\frac{2\mathrm{ln^2}(\delta)({N-1})}{w}} \geq \text{  } \frac{0.9}{N}
        . %\blacksquare
    \label{P2_app5} 
%\end{equation}
\end{align}

%\section{Other Solutions}
%\bibliographystyle{IEEEtran}
%\bibliography{bib}


\end{document}